\documentclass[twocolumn,epjc3]{svjour3}          

\usepackage[utf8]{inputenc}
 
\usepackage{amsmath,amssymb,amsbsy,graphicx,textcomp,enumerate,alltt,xspace,multirow,xcolor}
\usepackage{fancyvrb}
\usepackage[english]{babel}
\usepackage{subcaption}
\usepackage{float}
\usepackage{listings}

\newcommand\POWHEGBOX{{\tt POWHEG\,BOX}}
\newcommand\POWHEG{{\tt POWHEG}}

\newcommand\Pythia{{\tt Pythia}}
\newcommand\Herwig{{\tt Herwig}}

\newcommand\HerwigSeven{{\tt Herwig7}}
\newcommand\HerwigSevenPone{{\tt Herwig7.1}}

\newcommand\PythiaEightPtwo{{\tt Pythia8.2}}

\newcommand\PythiaSixPfour{{\tt Pythia6.4}}
\newcommand\PythiaSixPFour{{\tt Pythia6.4}}

\newcommand\HerwigSixPfive{{\tt Herwig6.5}}
\newcommand\HerwigSixPFive{{\tt Herwig6.5}}

\newcommand\sss{\mathchoice%
{\displaystyle}%
{\scriptstyle}%
{\scriptscriptstyle}%
{\scriptscriptstyle}%
}
\newcommand{\pt}{\ensuremath{p_{\sss\rm T}}\xspace}

\def\beq{\begin{equation}}
\def\beqn{\begin{eqnarray}}
\def\eeq{\end{equation}}
\def\eeqn{\end{eqnarray}}

\def\({\left(} 
\def\){\right)} 
 
\newcommand     \MSB            {\ifmmode {\overline{\rm MS}} \else
                                 $\overline{\rm MS}$\fi}

\newcommand\mt{m_{t}}

\newcommand\bjet{\ensuremath{b}-jet}

\newcommand\mwbj{\ensuremath{m_{Wb_j}}}
\newcommand\mwbjmax{\ensuremath{m_{Wb_j}^{\max}}}
\newcommand\Ebj{\ensuremath{E_{b_j}}}
\newcommand\Ebjmax{\ensuremath{E_{b_j}^{\max}}}
\newcommand\pT{\ensuremath{p_{\sss\rm  T}}}

\newcount\minutes 
\newcount\scratch 
\def\timestamp{%
\scratch=\time 
\divide\scratch by 60 
\edef\hours{\the\scratch} 
\multiply\scratch by 60 
\minutes=\time 
\advance\minutes by -\scratch 
---$\,$\hours:\null 
\ifnum\minutes< 10 0\fi 
\the\minutes}

\definecolor{mygray}{gray}{0.5}

\newcommand\PythiaEightPlot{{\tt Py8.2}}
\newcommand\HerwigSevenPlot{{\tt Hw7.1}}
\newcommand\HerwigSixPlot{{\tt Hw6.5}}
\newcommand\PythiaSixPlot{{\tt Py6.4}}

\newcommand\ttbnlodecPlot{$t\bar{t}dec$}
\newcommand\bbllllPlot{$b\bar{b}4\ell$}
\newcommand\hvqPlot{$hvq$}
\newcommand\FSR{{\tt FSR}}
\newcommand\SR{{\tt SR}}

\newcommand\hvq{\hvqPlot}
\newcommand\bbfourl{\bbllllPlot}
\newcommand\ttNLOdec{\ttbnlodecPlot}
\newcommand\ttbnlodec{\ttNLOdec}

\newcommand\cpp{c++}

\RequirePackage[T1]{fontenc}

\smartqed  

\RequirePackage{graphicx}
\RequirePackage{mathptmx}      
\RequirePackage{flushend}
\RequirePackage[numbers,sort&compress]{natbib}
\RequirePackage[colorlinks,citecolor=blue,urlcolor=blue,linkcolor=blue]{hyperref}

\journalname{Eur. Phys. J. C}

\usepackage{listings}
\newlength{\wfigsing}
\newlength{\wfigsingmulti}
\newlength{\wfigdoub}
\newlength{\wtablarge}

\def\whichjournal{0}
\ifnum \whichjournal=0 
  \wtablarge=\textwidth

\else
\fi

\begin{document}

\title{Addendum to: A Theoretical Study of Top-Mass Measurements at the LHC
  Using NLO+PS Generators of Increasing Accuracy}

\author{
  Silvia Ferrario Ravasio\thanksref{e1,addr1}
  \and
  Tom\'a\v{s} Je\v{z}o\thanksref{e2,addr2}
  \and
  Paolo Nason\thanksref{e3,addr3}
  \and
  Carlo Oleari\thanksref{e4,addr3}
}
\thankstext{e1}{e-mail: silvia.ferrario-ravasio@durham.ac.uk}
\thankstext{e2}{e-mail: tomas.jezo@physik.uzh.ch}
\thankstext{e3}{e-mail: paolo.nason@mib.infn.it}
\thankstext{e4}{e-mail: carlo.oleari@mib.infn.it}

\institute{IPPP, Department of Physics, Durham University, Durham, UK\label{addr1}
  \and
  Physics Institute, Universit\"at Z\"urich, Z\"urich, Switzerland\label{addr2}
  \and
  Universit\`a di Milano-Bicocca and INFN, Sezione di Milano-Bicocca,
  Piazza della Scienza 3, 20126 Milano, Italy\label{addr3}
}

\hyphenation{Mon-te}
\hyphenation{im-ple-ment-ed}

\maketitle

\begin{abstract}
  This paper is a follow-up of Ref.~\cite{Ravasio:2018lzi}, where we studied
  the impact of next-to-leading order calculations merged with parton shower
  generators (NLO+PS) of increasing accuracy in the extraction of the top
  mass at hadron colliders.  Here we examined results obtained with the older
  (fortran-based) shower generators \PythiaSixPFour{} and \HerwigSixPFive{}.
  Our findings are in line with what we found in Ref.~\cite{Ravasio:2018lzi}
  with the new, \cpp{}-based, generators \PythiaEightPtwo{} and
  \HerwigSevenPone{}.
\end{abstract}

\section{Introduction}
In Ref.~\cite{Ravasio:2018lzi} we considered three NLO+PS generators for
$t\bar{t}$ production, \hvq{}~\cite{Frixione:2007nw},
\ttbnlodec{}~\cite{Campbell:2014kua}, and \bbfourl{}~\cite{Jezo:2016ujg},
implemented in the \POWHEGBOX{}~\cite{Nason:2004rx, Frixione:2007vw,
  Alioli:2010xd, Jezo:2015aia}, interfaced with either \PythiaEightPtwo{}
(\PythiaEightPlot)~\cite{Sjostrand:2014zea} or \HerwigSevenPone{}
(\HerwigSevenPlot{})~\cite{Bahr:2008pv, Bellm:2015jjp}.  We focused
particularly on an observable that mimics those used in direct top mass
measurements, but also included in our study the proposed top mass
measurements from the peak energy of the $b$ jet~\cite{Agashe:2016bok} and
from the class of leptonic observables suggested in
Ref.~\cite{Frixione:2014ala}.  We found large differences between predictions
obtained using the two parton shower programs. In particular, while results
obtained with the three NLO+PS generators interfaced to \PythiaEightPlot{}
are fairly consistent among each other, large differences are found if they
are interfaced to \HerwigSevenPlot{}.

In this addendum we discuss the results obtained with the older,
fortran-based versions of the \Pythia{} and \Herwig{} codes.  Our purpose is
to see if the effects that we have seen are specific to the new
implementations, or were already present in the old ones.  We briefly recall
the characteristics of the older generators:
\begin{itemize}
\item \PythiaSixPfour{}~(\PythiaSixPlot{})~\cite{Sjostrand:2006za}:
  implements a \pT-ordered shower, making use of the same algorithm adopted
  in \PythiaEightPlot{}.  The older and new codes have both an interleaved
  radiation scheme between the initial-state radiation and the multi-parton
  interactions~(MPI).  In \PythiaEightPlot{}, final-state radiation
  is also interleaved, and different models of colour
  reconnection are also offered.
  
\item \HerwigSixPfive{}~\cite{Corcella:2000bw} with {\tt
  Jimmy\,4.31}~\cite{Butterworth:1996zw}~(\HerwigSixPlot{}): implements an
  angular-ordered shower.  However, the showering variables are different
  from those adopted in the \HerwigSevenPlot{}
  implementation~\cite{Gieseke:2003rz}.
  The two versions of \Herwig{} implement the PS and the perturbative part of
  the MPI in a similar manner.  The non-perturbative part of the MPI,
  instead, has been completely redesigned~\cite{Bahr:2008dy}.
  Similarly to \Pythia{}, colour-reconnection effects are properly included
  only in the recent versions of \Herwig{}~\cite{Gieseke:2012ft}.
\end{itemize}
In our previous work, we have seen that the two generators \bbfourl{} and
\ttbnlodec{} yield fairly consistent results for the observables that we have
considered. Thus, here we only compare \hvq{} and \bbfourl{}.

\section{Interface to \POWHEGBOX{}}
In this section we briefly describe the matching of \bbfourl{} and \hvq{} to
both \PythiaSixPlot{} and \HerwigSixPlot{}.
The matching to \PythiaEightPlot{} and \HerwigSevenPlot{} is detailed in
Ref.~\cite{Ravasio:2018lzi}.

\subsection{\PythiaSixPfour{}}
\PythiaSixPlot{} implements both a $\pT$ and a virtuality-ordered PS.  Here,
we employ the $\pT$-ordered shower with the Perugia
tune~({\tt PYTUNE(320)})~\cite{Skands:2010ak}.

We setup \PythiaSixPlot{} in such a way that the $\pt$ of radiation
in the shower is limited by the {\tt scalup} parameter of the
Les Houches Interface for User Processes~\cite{Boos:2001cv}, as is usually
done in \POWHEG{}. This is at variance
with the  Perugia tune settings, that requires $\pt$ to be smaller than
{\tt scalup} divided by $\sqrt{2}$.\footnote{We achieve this
  by setting the \PythiaSixPlot{} parameter {\tt PARP(71)=4} rather
  than the default Perugia value {\tt PARP(71)=2}.}

The matching of shower emissions in the production process relies on the
default behaviour of \POWHEG{}, i.e.~the shower evolution starts at {\tt
  scalup}.  In the decays, a different scale must be adopted, and thus it
requires a custom veto prescription in \bbfourl{}.  We implement it using two
methods, both analogous to what we did in order to match \PythiaEightPlot{}
to \bbfourl{} in Ref.~\cite{Ravasio:2018lzi}:

\begin{enumerate}
\item Each time \PythiaSixPfour{} generates an emission off the top (or
  anti-top), we compute its transverse momentum according to the \POWHEG{}
  definition.  If it is larger than the transverse momentum of the emission
  generated by the \POWHEGBOX{}, we abandon the current shower, and restart a
  shower from the same Les Houches event.  This represents our default
  method.  We label it as the ``{\tt FSR}'' veto, in full analogy with
  the notation adopted for \PythiaEightPlot{}.

\item Since we employ a $\pT$-ordered shower, we can also simply
  require the shower to start at a given transverse momentum,
  that we set equal to the transverse momentum of the corresponding
  \POWHEG{} emission.
  This veto procedure will be referred to as the ``{\tt SR}'' method, as we
  did with the analogous method that we adopted in \PythiaEightPlot{}.
\end{enumerate}

\subsection{\HerwigSixPfive{}}

For {\tt Hw6+Jimmy} we adopted the ATLAS AUET2 tune~\cite{ATLAS:2011gmi}.
The \Herwig{} shower is ordered in angle and not in $\pT$. Therefore all the
emissions with transverse momentum larger than that of the \POWHEG{}
emission must be vetoed. Both \Herwig{} versions already enforce this veto
for the production part of the process.  Similarly to \PythiaSixPlot{}, extra
care is required for emissions from the top-decay products, when interfaced
with \bbfourl{}.

In our previous work, two procedures were devised to veto extra
\HerwigSevenPlot{} emissions.  Both of them use the \pT{} of the \POWHEG{}
emission as an upper bound, either on the \pT{} of each branching at the end
of the showering phase ({\tt FullShowerVeto}), or on the shower evolution
scale during the showering phase ({\tt ShowerVeto}).  Unfortunately, the
\HerwigSixPlot{} event record (as for \PythiaSixPlot{}) does not contain
information regarding the branching of the partons, i.e.~it is not possible
to reconstruct the emission's history after the shower is completed, in
contrast to the new version of the code.  Therefore, we only implemented the
analogue of the \HerwigSevenPlot{} {\tt ShowerVeto} method which proceeds as
follows: when an emission off a top resonance is generated, if its $\pT$
(defined in terms of \Herwig{} variables) is larger than that of the
\POWHEG{} emission, the branching is discarded and the evolution continues
from the scale of this discarded emission.

\section{Hadronic observables: NLO+PS results}
In this section we compare predictions for hadronic observables at the NLO+PS
level, i.e.~without the inclusion of MPI
and of hadronization effects.  Our
aim is to assess differences of perturbative origin and, in particular, due
to the NLO+PS matching.

\subsection{\PythiaSixPfour{} versus \PythiaEightPtwo{}}
We begin by comparing the predictions obtained with \PythiaSixPlot{} and
\PythiaEightPlot{}, which both implement a dipole-like algorithm for
final-state showers.

\begin{figure} 
  \begin{center}
    \hspace{0.6em}
    \includegraphics[width=0.46\textwidth]{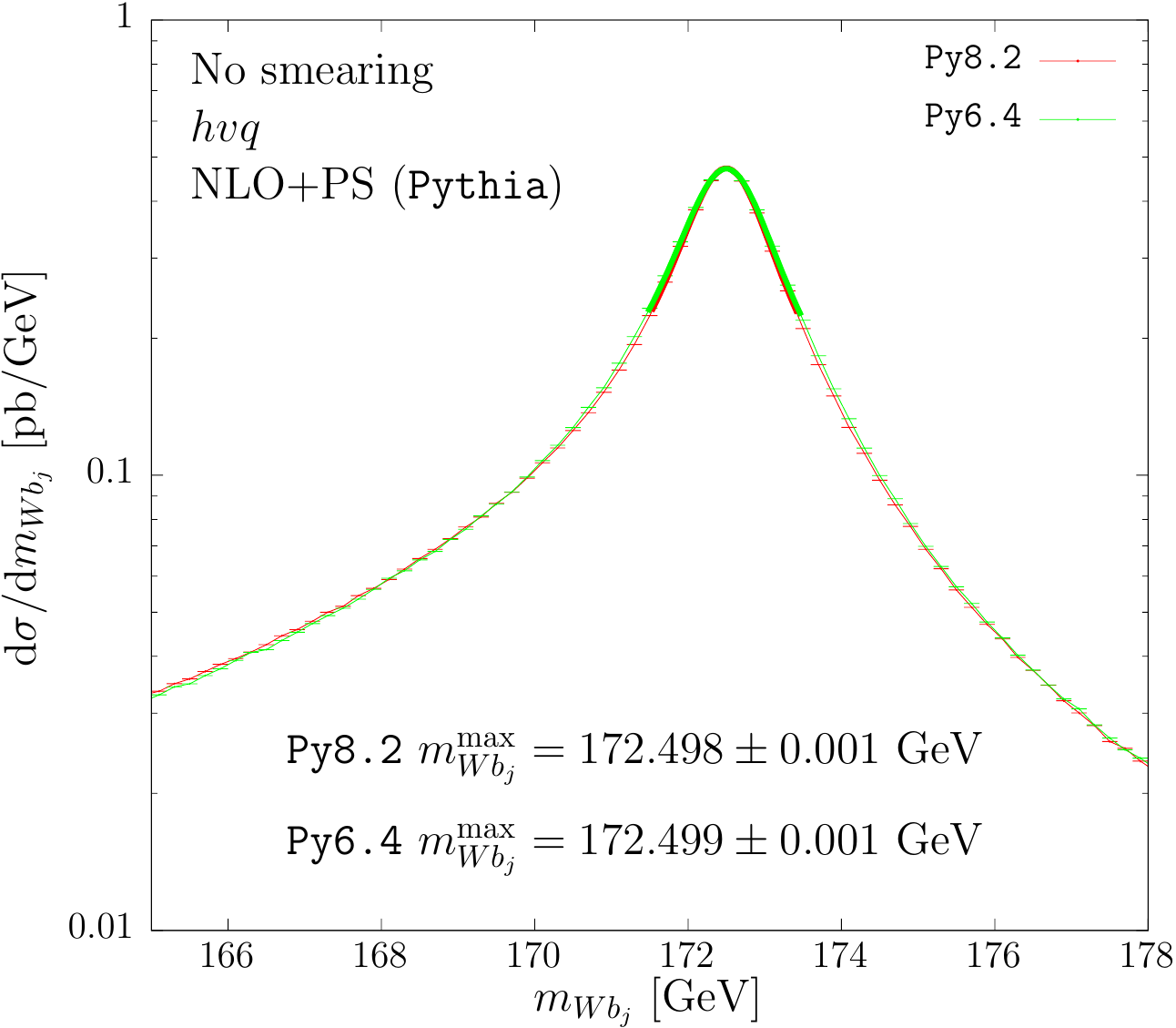}\\
    \includegraphics[width=0.47\textwidth]{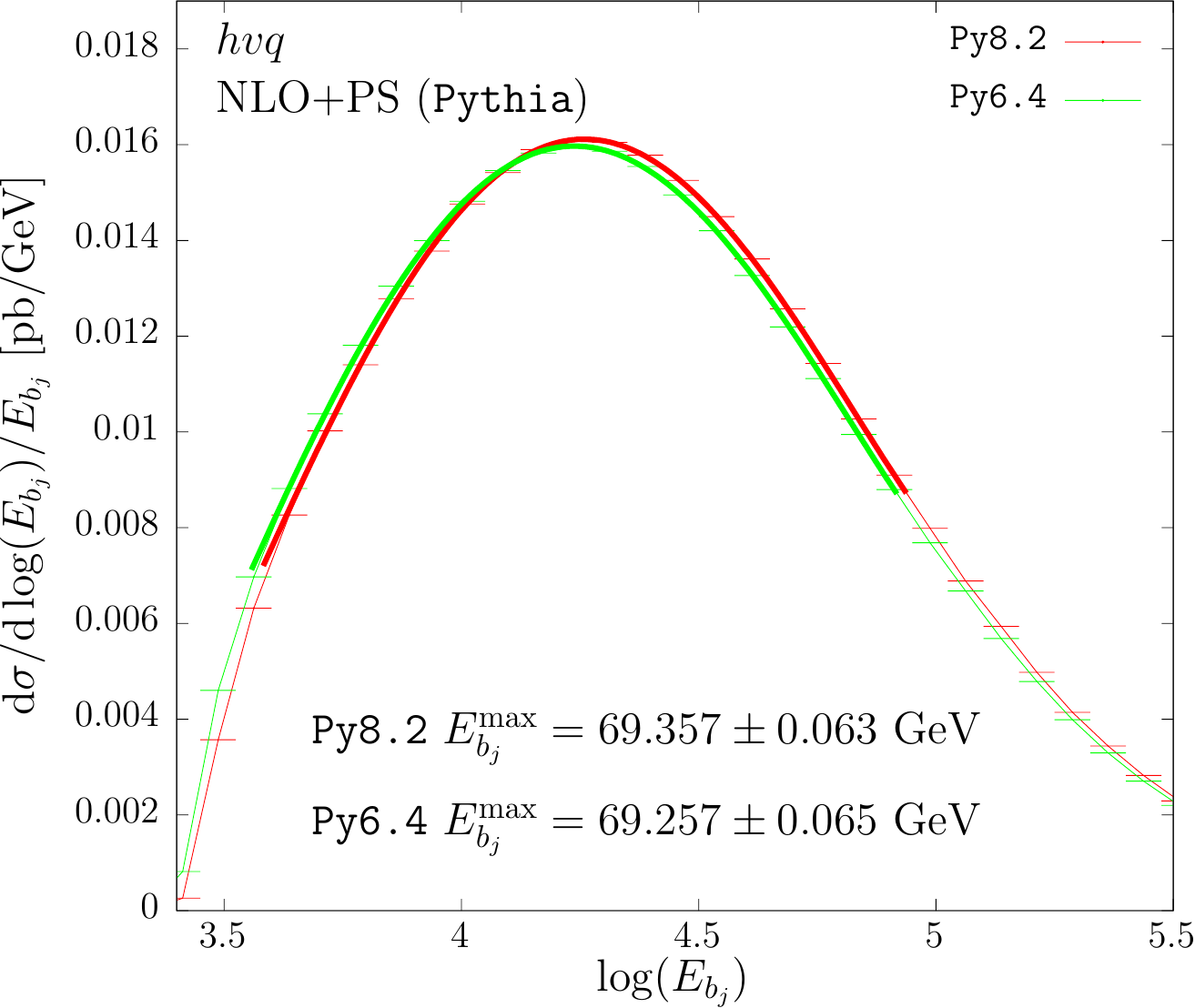}
  \end{center}
  \caption{Reconstructed-top mass~(upper pane) and \bjet{} energy
    distribution~(lower pane) obtained with the \hvq{} generator interfaced
    to \PythiaEightPlot{}~(red) and to \PythiaSixPlot{}~(green).
    Hadronization and MPI effects are not included.}
  \label{fig:hvq_py_showerOnly}
\end{figure}

In Ref.~\cite{Ravasio:2018lzi} we made use of a smearing procedure to simulate experimental
resolution effects. We begin by examining results obtained without applying
any smearing.

The distributions of the reconstructed-top mass and of the \bjet{} energy
using \hvq{} matched to the two versions of \Pythia{} are shown in the upper
and lower panes of Fig.~\ref{fig:hvq_py_showerOnly}, respectively.  The two
curves for the reconstructed-top mass are almost indistinguishable.  Also the
peak positions of the \bjet{} energy spectra agree remarkably well, despite
some small differences in shape, leading to a displacement of the extracted
top-mass for this observable of $\approx 200$~MeV.

\begin{figure} 
  \begin{center}
    \hspace{0.2em}
    \includegraphics[width=0.46\textwidth]{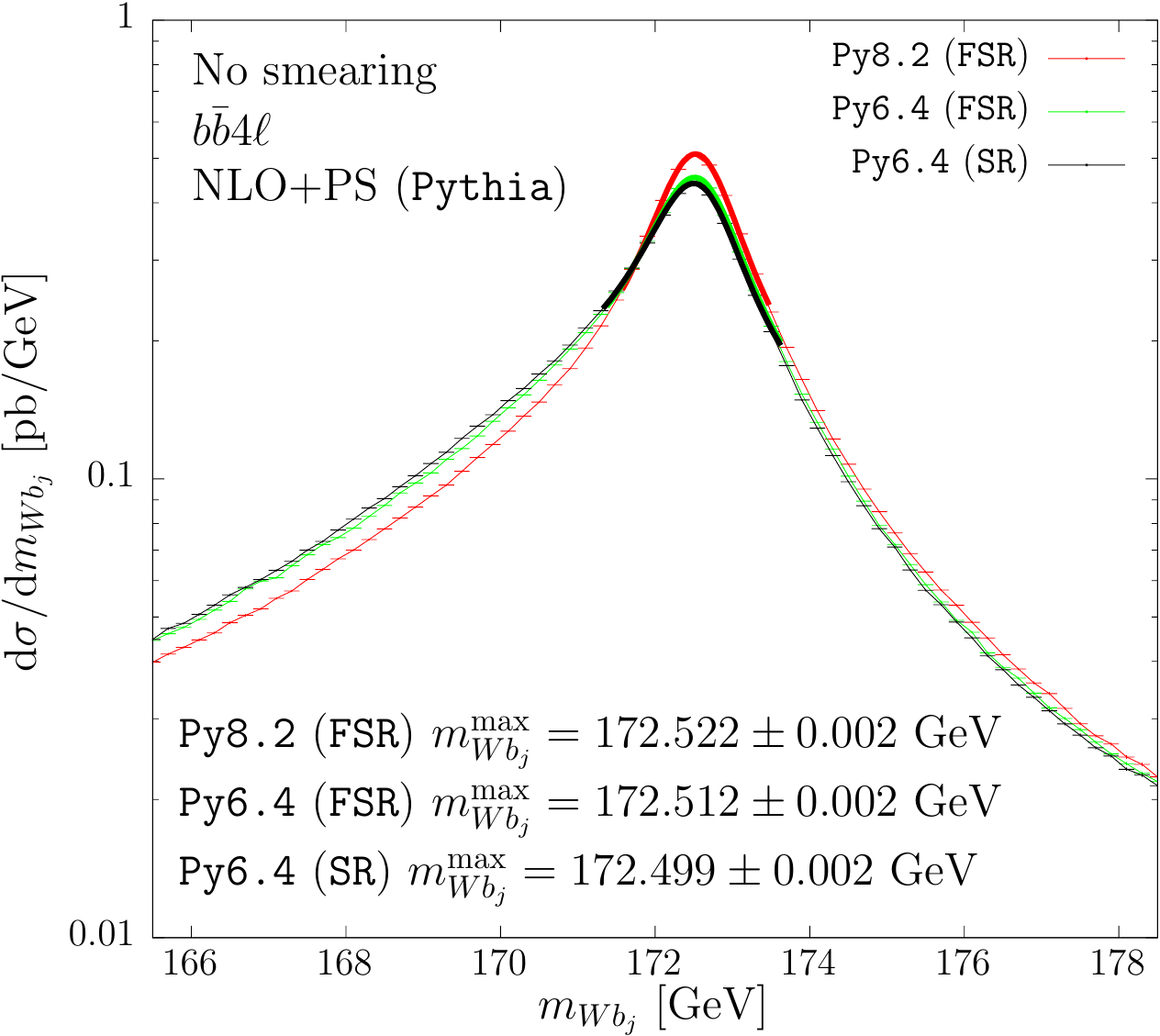}\\
    \includegraphics[width=0.475\textwidth]{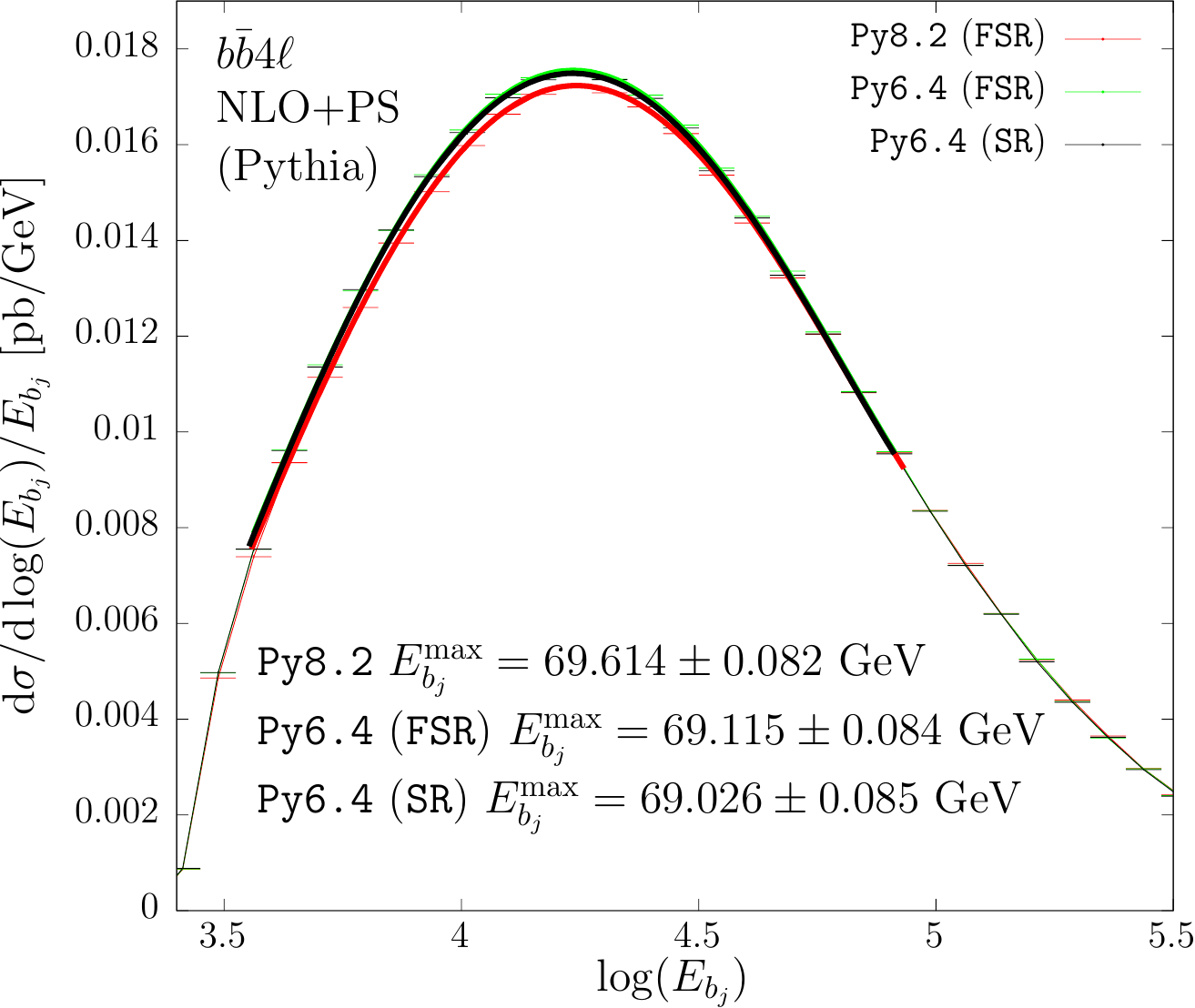}
  \end{center}
  \caption{Reconstructed-top mass~(upper pane) and \bjet{} energy
    distributions~(lower pane) obtained with the \bbfourl{} generator
    showered by \PythiaEightPlot{} with the {\tt FSR} veto scheme~(red), and
    by \PythiaSixPlot{}. The two curves for the \PythiaSixPlot{} results are
    obtained using the {\tt FSR} veto scheme~(green) and the {\tt SR} veto
    scheme~(black). Hadronization and MPI effects are not included.}
  \label{fig:bb4l_py_showerOnly}
\end{figure}
In Fig.~\ref{fig:bb4l_py_showerOnly} we plot the distributions obtained using
the \bbfourl{} generator. The results for the \mwbj{} spectrum obtained with
\PythiaSixPlot{} show an enhancement in the low-mass region with respect to
the \PythiaEightPlot{} distribution, irrespective of the veto scheme used
(upper pane). Nevertheless there is no appreciable shift in the
peak-position.

The shape of the \bjet{} energy spectrum in the proximity of the peak region
is instead different for \PythiaEightPlot{} compared to the two results
obtained by using \PythiaSixPlot{}, with a shift in the maximum of the
\bjet{} energy of approximately +0.5~GeV of the former with respect to the
latter two results. This shift induces a displacement in the extracted
top-mass~($m_t$) of $\approx 1$~GeV.\footnote{See eqs.~(7.2) and~(7.4) of
  Ref.~\cite{Ravasio:2018lzi}.}

\begin{table*} 
\resizebox{1.\textwidth}{!}  { \begin{tabular}{l|c|c|c|c|c|c|}
 \cline{2-7}
 &  \multicolumn{2}{ |c|}{ \phantom{\Big|} $R=0.4$}
 &  \multicolumn{2}{ |c|}{ \phantom{\Big|} $R=0.5$}
 &  \multicolumn{2}{ |c|}{ \phantom{\Big|} $R=0.6$} \\
 \cline{2-7}
 & \phantom{\Big|} No smearing & 15~GeV smearing
 & \phantom{\Big|} No smearing & 15~GeV smearing
 & \phantom{\Big|} No smearing & 15~GeV smearing \\
 \cline{1-7}
 \multicolumn{1}{ |c|  }{ \phantom{\Big|}\bbfourl{}+\PythiaEightPlot{} (\FSR{})~[GeV]}

 & $ 172.509\pm  0.002$ & $ 170.569\pm  0.002$

 & $ 172.522\pm  0.002$ & $ 171.403\pm  0.002$

 & $ 172.538\pm  0.002$ & $ 172.117\pm  0.002$
\\ \cline{1-7}
\multicolumn{1}{ |c|  }{ \phantom{\Big|}\bbfourl{}+\PythiaSixPlot{} (\FSR{}) ${}-$ \bbfourl{}+\PythiaEightPlot{} (\FSR{})}
 & $     -22
\pm      3$~MeV
 & $    -296
\pm      2$~MeV
 & $     -11
\pm      3$~MeV
 & $    -286
\pm      2$~MeV
 & $       0
\pm      3$~MeV
 & $    -258
\pm      2$~MeV
\\ \cline{1-7}
\multicolumn{1}{ |c|  }{ \phantom{\Big|}\bbfourl{}+\PythiaSixPlot{} (\SR{}) ${}-$ \bbfourl{}+\PythiaEightPlot{} (\FSR{})}
 & $     -36
\pm      3$~MeV
 & $    -360
\pm      2$~MeV
 & $     -23
\pm      3$~MeV
 & $    -342
\pm      2$~MeV
 & $      -8
\pm      3$~MeV
 & $    -307
\pm      2$~MeV
\\ \cline{1-7}
 \\[-1.25em]
\cline{1-7}
 \multicolumn{1}{ |c|  }{ \phantom{\Big|}\hvq{}+\PythiaEightPlot{}~[GeV]}

 & $ 172.485\pm  0.001$ & $ 170.518\pm  0.001$

 & $ 172.498\pm  0.001$ & $ 171.315\pm  0.001$

 & $ 172.513\pm  0.001$ & $ 171.996\pm  0.001$
\\ \cline{1-7}
\multicolumn{1}{ |c|  }{ \phantom{\Big|}\hvq{}+\PythiaSixPlot{} ${}-$ \hvq{}+\PythiaEightPlot{}}
 & $     -11
\pm      2$~MeV
 & $ +    76
\pm      2$~MeV
 & $ +     1
\pm      2$~MeV
 & $ +    69
\pm      2$~MeV
 & $ +    13
\pm      2$~MeV
 & $ +    69
\pm      2$~MeV
\\ \cline{1-7}
\end{tabular}
}
\caption{Comparisons between the \PythiaEightPlot{} and the \PythiaSixPlot{}
  results for \mwbjmax{}, computed with \bbfourl{} and \hvq{}, without
  hadronization or MPI effects, for different values of the jet radius $R$. }
     \label{tab:mwbj_py_showerOnly}  
\end{table*}
\begin{table*} 
  \centering
  { \begin{tabular}{l|c|c|c|}
 \cline{2-4}
 &  \multicolumn{1}{ |c|}{ \phantom{\Big|} $R=0.4$}
 &  \multicolumn{1}{ |c|}{ \phantom{\Big|} $R=0.5$}
 &  \multicolumn{1}{ |c|}{ \phantom{\Big|} $R=0.6$} \\
 \cline{1-4}
 \multicolumn{1}{ |c|  }{ \phantom{\Big|}\bbfourl{}+\PythiaEightPlot{} (\FSR{})~[GeV]}

 & $  67.145\pm  0.086$

 & $  69.614\pm  0.082$

 & $  71.747\pm  0.080$
\\ \cline{1-4}
\multicolumn{1}{ |c|  }{ \phantom{\Big|}\bbfourl{}+\PythiaSixPlot{} (\FSR{}) ${}-$ \bbfourl{}+\PythiaEightPlot{} (\FSR{})}
 & $    -422
\pm    124$~MeV
 & $    -499
\pm    118$~MeV
 & $    -512
\pm    115$~MeV
\\ \cline{1-4}
\multicolumn{1}{ |c|  }{ \phantom{\Big|}\bbfourl{}+\PythiaSixPlot{} (\SR{}) ${}-$ \bbfourl{}+\PythiaEightPlot{} (\FSR{})}
 & $    -455
\pm    123$~MeV
 & $    -588
\pm    118$~MeV
 & $    -543
\pm    114$~MeV
\\ \cline{1-4}
 \\[-1.25em]
\cline{1-4}
 \multicolumn{1}{ |c|  }{ \phantom{\Big|}\hvq{}+\PythiaEightPlot{}~[GeV]}

 & $  66.791\pm  0.068$

 & $  69.357\pm  0.063$

 & $  71.598\pm  0.061$
\\ \cline{1-4}
\multicolumn{1}{ |c|  }{ \phantom{\Big|}\hvq{}+\PythiaSixPlot{} ${}-$ \hvq{}+\PythiaEightPlot{}}
 & $     -24
\pm     95$~MeV
 & $    -100
\pm     91$~MeV
 & $    -133
\pm     87$~MeV
\\ \cline{1-4}
\end{tabular}
}
\caption{Comparisons between the \PythiaEightPlot{} and the \PythiaSixPlot{}
  results for \Ebjmax{}, computed with \bbfourl{} and \hvq{}, without
  hadronization or MPI effects, for different values of the jet radius $R$.}
    \label{tab:Ebj_py_showerOnly} 
\end{table*}

In Tabs.~\ref{tab:mwbj_py_showerOnly} and~\ref{tab:Ebj_py_showerOnly} we
summarize the \mwbj{} and \Ebj{} peak positions respectively, obtained for
different values of the jet radius varied between 0.4 and 0.6.
Table~\ref{tab:mwbj_py_showerOnly} also shows the \mwbj{} distribution peak
positions when the smearing is applied.  An excellent agreement is found
between \hvq{}+\PythiaSixPlot{} and \hvq{}+\PythiaEightPlot{} for \mwbjmax{},
even after the smearing is applied, and the \Ebjmax{} differences are small, nearly
consistent with zero within their statistical errors for all values of $R$.

The low-mass enhancement in the \mwbj{} spectrum of the
\bbfourl{}+\PythiaSixPlot{} generator, with respect to the
\bbfourl{}+\PythiaEightPlot{} generator, leads to quite large displacements
of the peak position once smearing is applied.  For our default {\tt
  FSR}-veto procedure, the differences between \PythiaEightPlot{} and
\PythiaSixPlot{} are roughly 250-300~MeV.  The differences of \Ebjmax{} for
the two showers used with \bbfourl{} are even larger, of the order of 0.5~GeV
for all values of the jet radius.

The differences in \mwbjmax{} and \Ebjmax{} between the \bbfourl{} and \hvq{}
generators for $R=0.5$ are reported in Tab.~\ref{tab:py_showerOnly_diff}.
\begin{table}[h]
\centering
\begin{tabular}{c|c|c|c|}
\cline{2-4}
& \multicolumn{3}{|c|}{\phantom{\Big|}\bbfourl{}${}-$\hvq{}, \; $R=0.5$\qquad [MeV] \phantom{\Big|}} \\
\cline{2-4}
\phantom{\Big|}& \mwbjmax & \mwbjmax{} (smear) & \Ebjmax \\
\cline{1-4}
\multicolumn{1}{|c|}{\phantom{\Big|} \PythiaEightPlot{} (\FSR) \phantom{\Big|}}
 & $24 \pm           2$ & $89 \pm           2$ & $257 \pm 53$
 \\ \cline{1-4}
\multicolumn{1}{|c|}{\phantom{\Big|} \PythiaSixPlot{} (\FSR) \phantom{\Big|}}
 & $12 \pm           2 $ & $-265 \pm           2$ & $-147 \pm 106$
\\ \cline{1-4}
\end{tabular}
\caption{Differences between the \bbfourl{} and \hvq{} predictions for
  \mwbjmax{} (with and without smearing) and \Ebjmax{}, showered by
  \PythiaEightPlot{} and \PythiaSixPlot{}.}
\label{tab:py_showerOnly_diff}
\end{table}
We notice that the level of agreement of
\mwbjmax{} predictions obtained using \bbfourl{} and \hvq{} gets worse in
\PythiaSixPlot{} as compared to \PythiaEightPlot{}, while the opposite is
true for \Ebjmax{}.

\subsection{\HerwigSixPfive{} versus \HerwigSevenPone{}}

We now compare the predictions obtained by showering the NLO+PS results with
\HerwigSixPlot{} and \HerwigSevenPlot{}.

\begin{figure} 
  \begin{center}
  \hspace{0.5em}\includegraphics[width=0.47\textwidth]{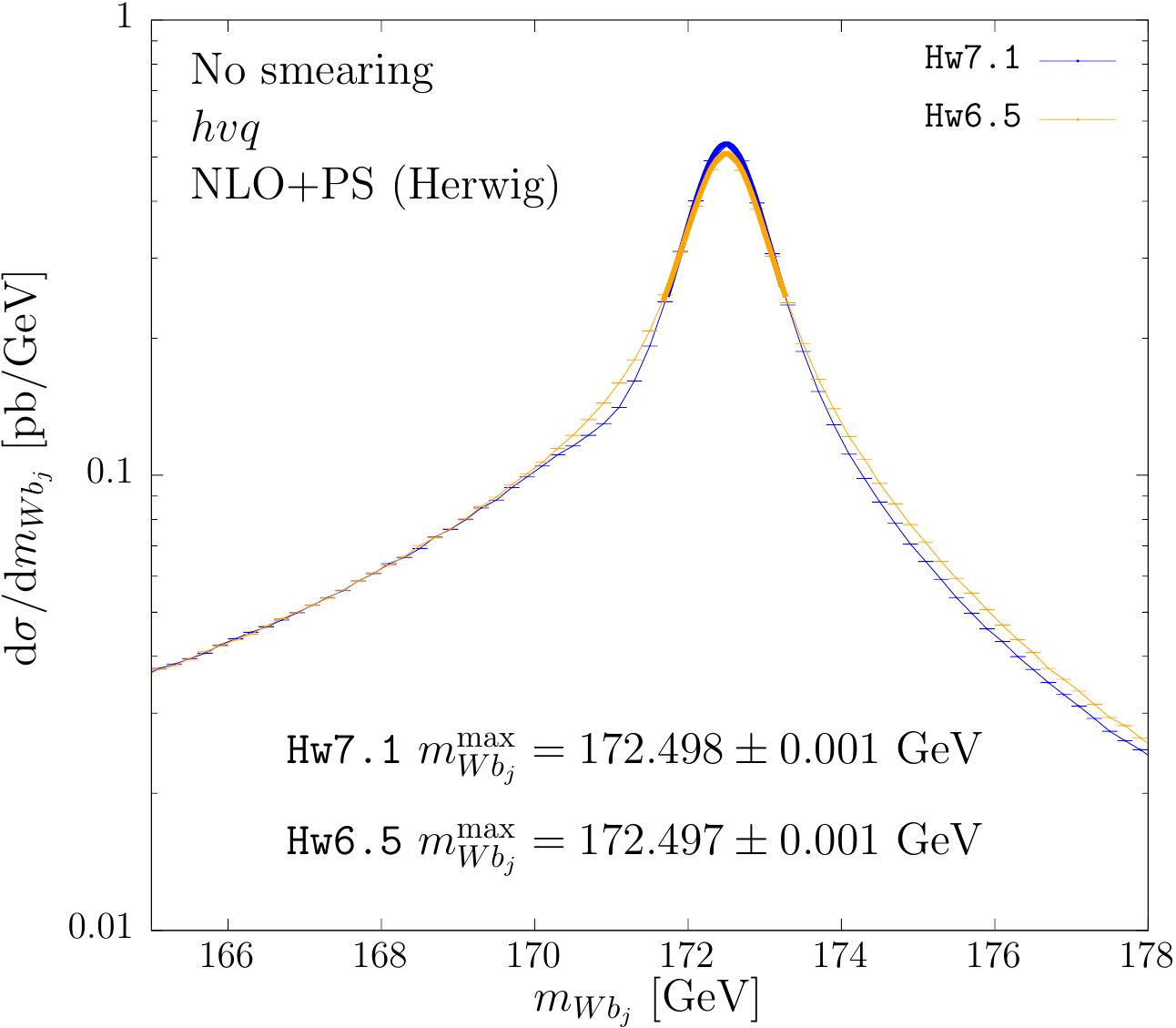}\\
  \includegraphics[width=0.48\textwidth]{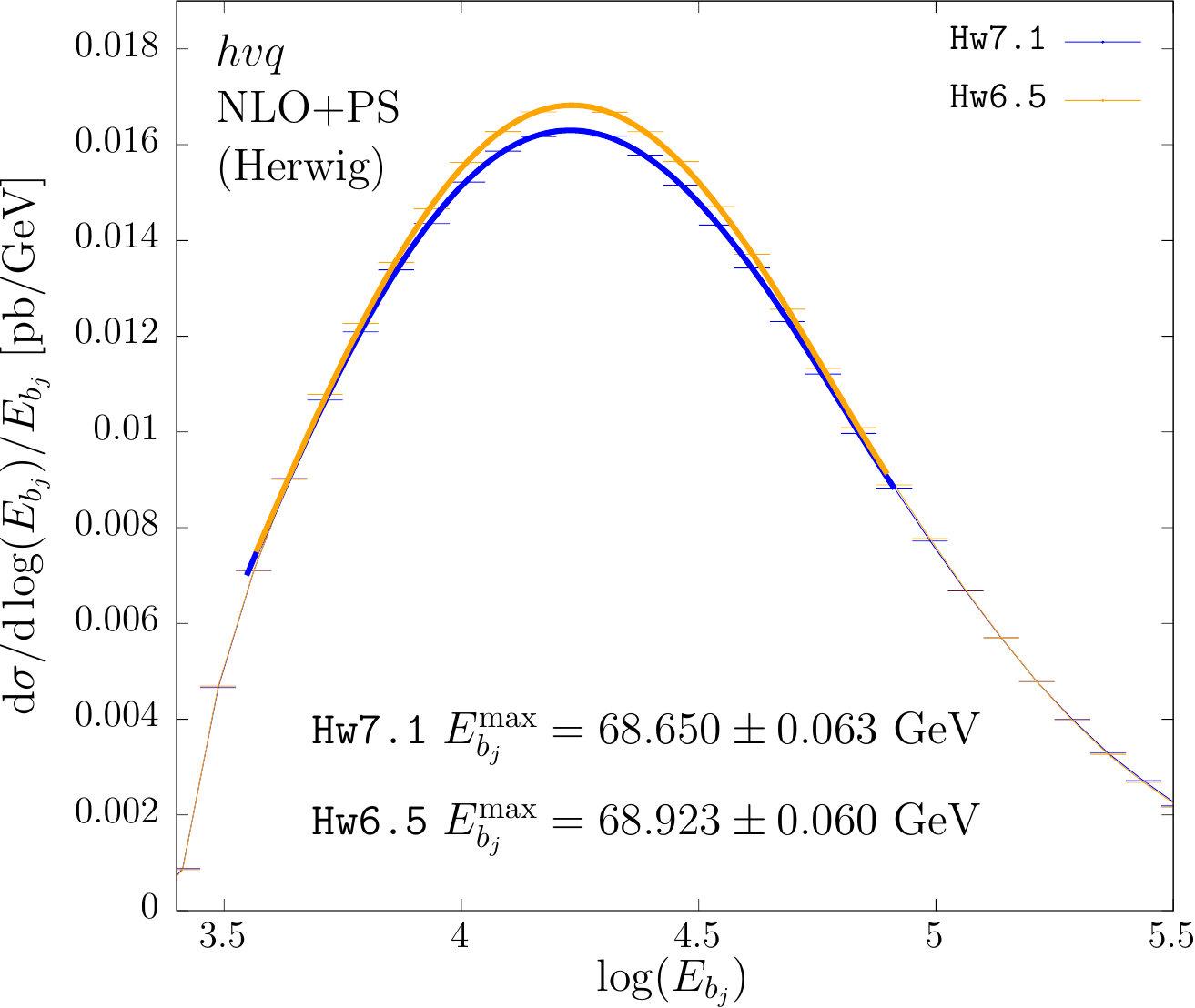}
  \end{center}
    \caption{Reconstructed-top mass~(upper pane) and $b$ jet energy
      distribution~(lower pane) computed with the \hvq{} generator matched to
      \HerwigSevenPlot{}~(blue) and to \HerwigSixPlot{}~(orange).
      Hadronization and MPI effects are not included.}
    \label{fig:hvq_hw_showerOnly}
\end{figure}

\begin{figure}
  \begin{center}
  \hspace{0.5em}\includegraphics[width=0.47\textwidth]{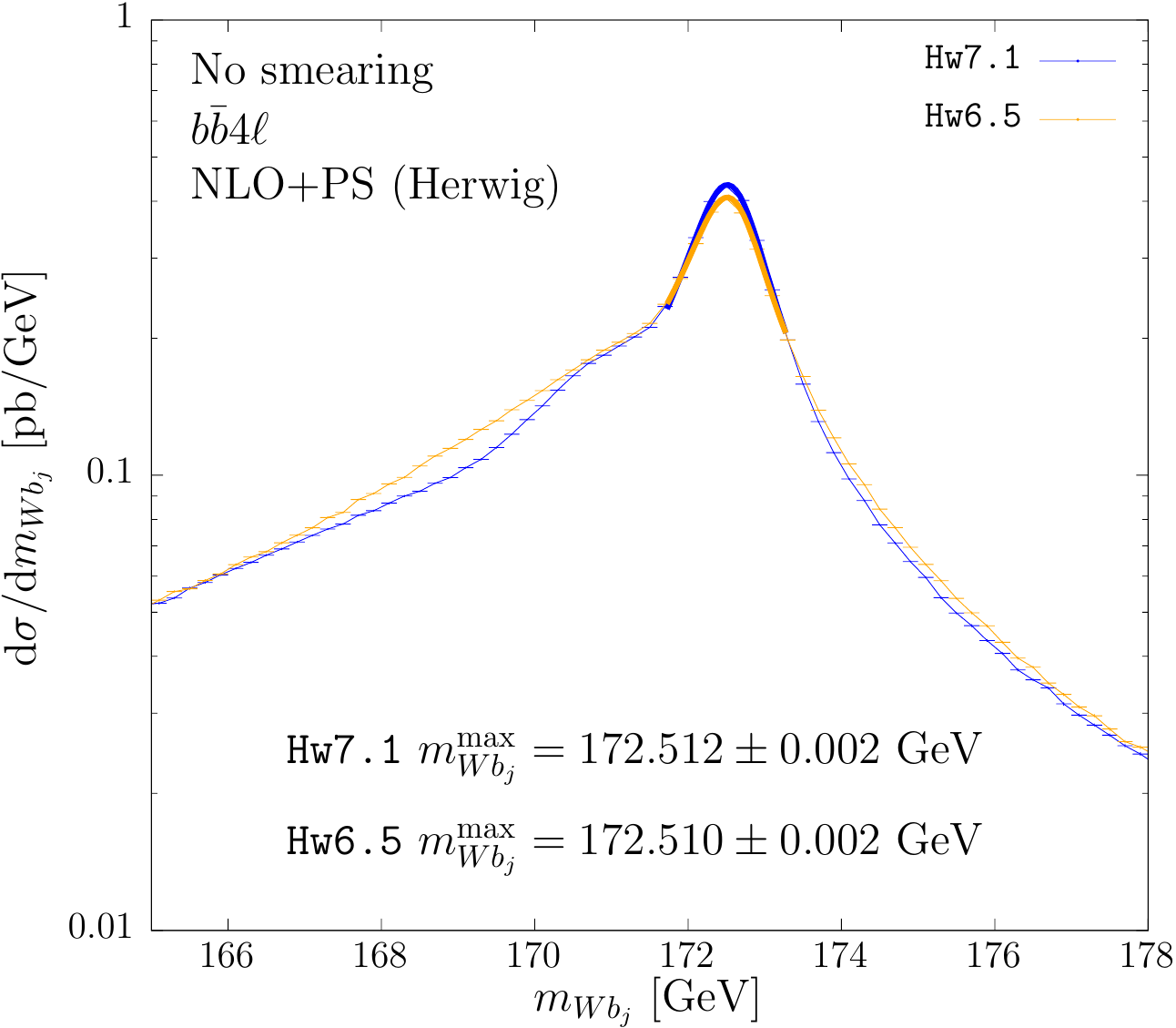}\\
  \includegraphics[width=0.48\textwidth]{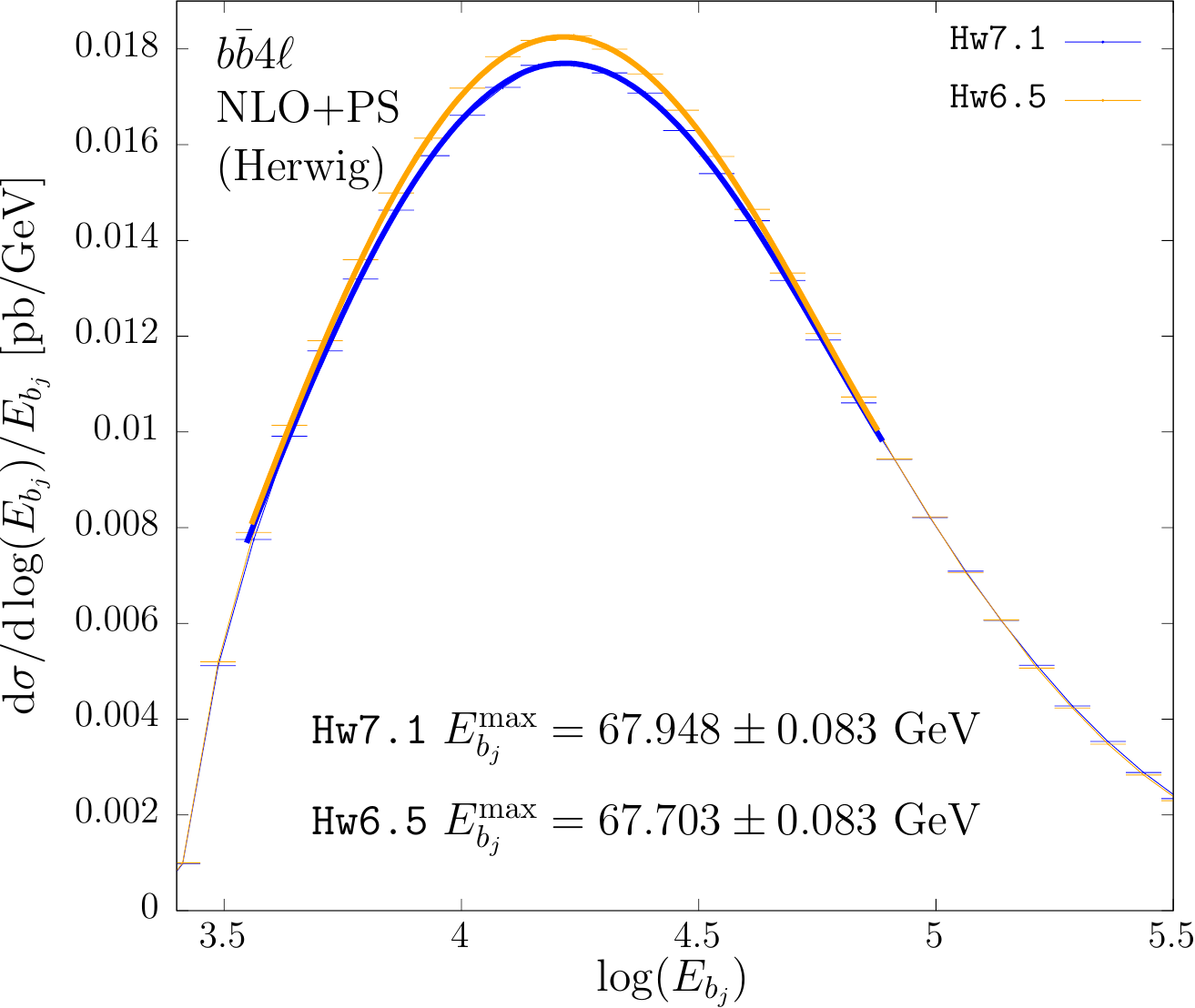}
  \end{center}
    \caption{Reconstructed-top mass~(upper pane) and $b$ jet energy
      distribution~(lower pane) computed with the \bbfourl{} generator
      matched to \HerwigSevenPlot{}~(blue) and to \HerwigSixPlot{}~(orange).
      Hadronization and MPI effects are not included.}
    \label{fig:bb4l_hw_showerOnly}
\end{figure}

In the upper panes of Figs.~\ref{fig:hvq_hw_showerOnly}
and~\ref{fig:bb4l_hw_showerOnly} we plot the results for \mwbj{} obtained
with \hvq{} and \bbfourl{}. The cross section under the peak is mildly
suppressed in \HerwigSixPlot{} with respect to \HerwigSevenPlot{}. This is
then compensated by enhancements in the low- and, to a smaller extent,
high-tail regions. A small bump is also present at roughly 1~GeV below the
peak position when using the \bbfourl{} generator with \HerwigSevenPlot{},
also present to a smaller extent when using \HerwigSixPlot{} instead.%
\footnote{Further studies suggest that this bump is a symptom of a minor
  shower cut-off mismatch between \HerwigSevenPlot{} and \bbfourl{}.} These
differences, present already at the shower level, could be ascribed to the
fact that the two versions of \Herwig{} adopt slightly different ordering
variables.%
\footnote{In \HerwigSixPlot{} the variable $z$ is interpreted as the energy
  fraction of the emitter after the emission, while in \HerwigSevenPlot{} it
  represents the light-cone momentum fraction.  In both, the ordering
  variable in the collinear limit becomes $\sim E \theta$, $E$ being the
  energy of the emitting parton and $\theta$ the angle between the two
  radiated partons.  See~\cite{Bahr:2008pv} for further details.}  Despite
the presence of these differences, the peak position (at the unsmeared level)
in \HerwigSixPlot{} or \HerwigSevenPlot{}, in both \hvq{} and
\bbfourl{}, is not changed.

In the lower panes of Figs.~\ref{fig:hvq_hw_showerOnly}
and~\ref{fig:bb4l_hw_showerOnly} we show the results for the \bjet{} energy
spectrum. The peak position, when \hvq{} is used, is 250~MeV
bigger when showering with \HerwigSixPlot{} than with
\HerwigSevenPlot{}, while in the case of \bbfourl{} it has the same magnitude
but opposite sign. This affects the extracted top mass by 0.5~GeV.

\subsection{\Pythia{} versus \Herwig{}}

\begin{figure}
  \centering
  \includegraphics[width=0.5\textwidth]{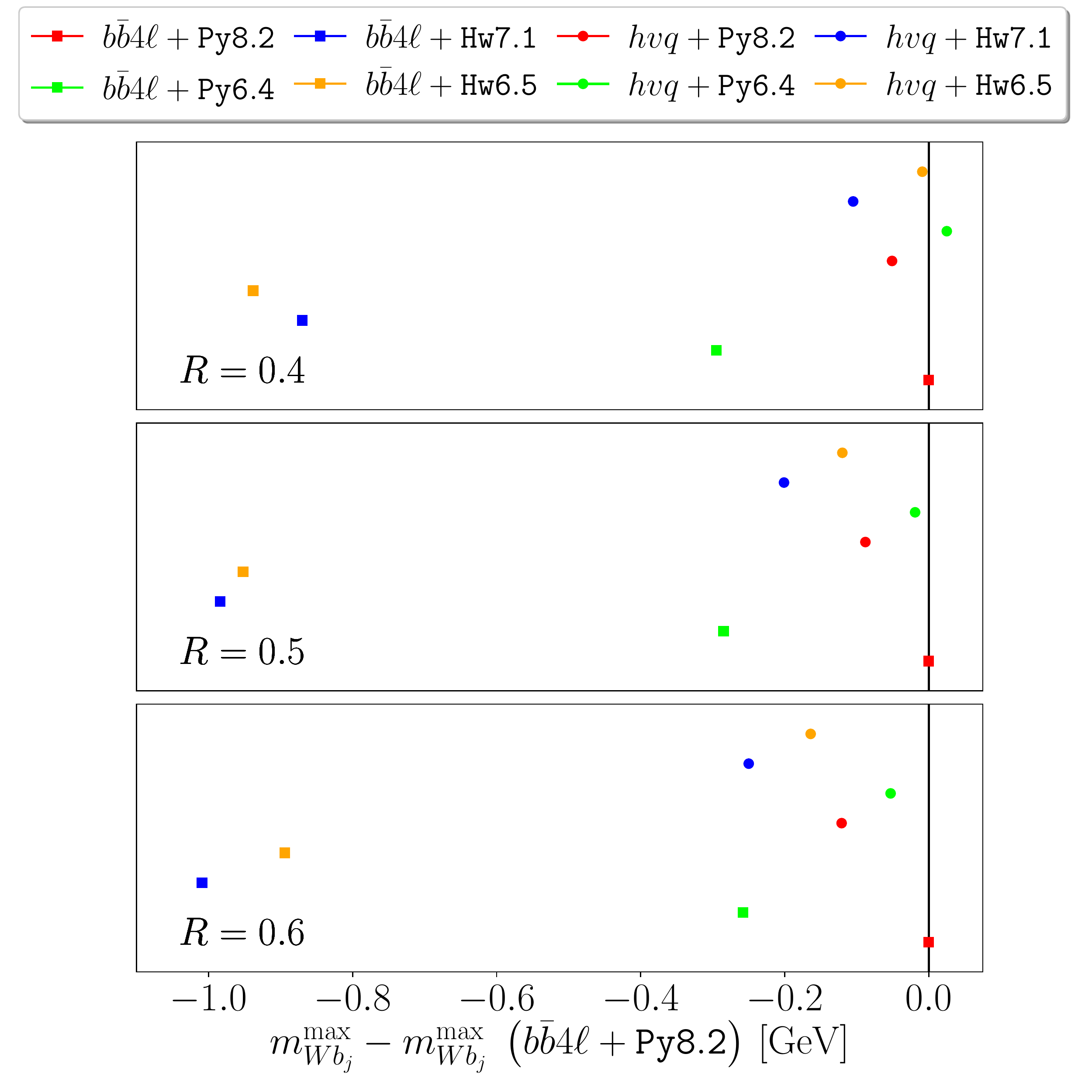}
  \caption{Results for the difference of the \mwbjmax{}, including a 15~GeV
    smearing, with respect to our reference generator
    (i.e.~\bbfourl+\PythiaEightPlot), at the NLO+PS level using \hvq{} or
    \bbfourl{}, showered by \Pythia{} and \Herwig{}, for different values of
    jet radius $R$.  Hadronization and MPI effects are not included.  The
    numerical values are reported in Tab.~\ref{tab:had_showerOnly}.}
  \label{fig:had_showerOnly_mwbj}
\end{figure}

\begin{figure}
  \centering
  \includegraphics[width=0.5\textwidth]{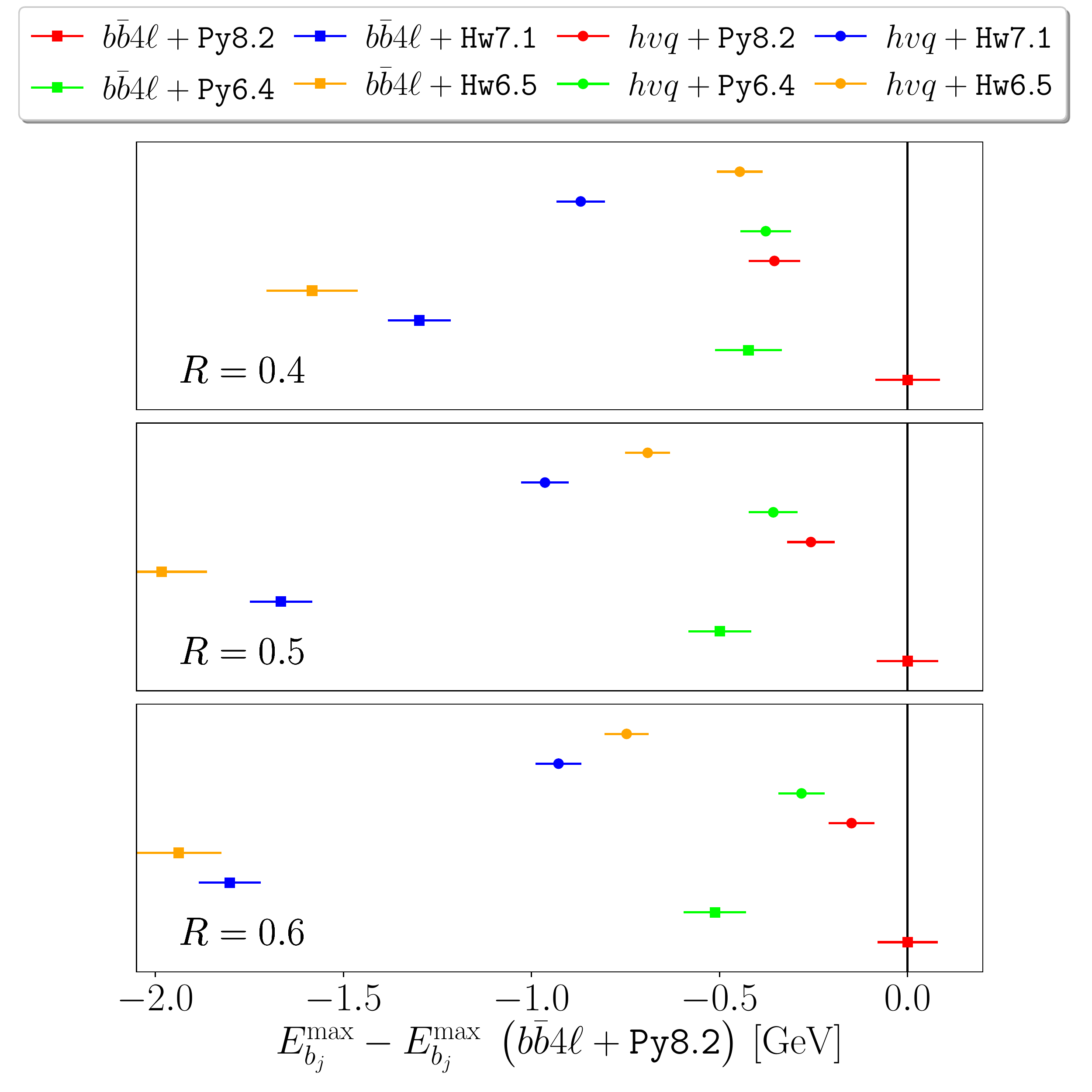}
  \caption{Same as Fig.~\ref{fig:had_showerOnly_mwbj} but for \Ebjmax.}
  \label{fig:had_showerOnly_Ebj}
\end{figure}

In Figs.~\ref{fig:had_showerOnly_mwbj} and~\ref{fig:had_showerOnly_Ebj} we
plot the variation of \mwbjmax{} and \Ebjmax{} (relative to our reference
generator combination, i.e.~\bbfourl+\PythiaEightPlot) obtained with
\bbfourl{} and \hvq{}, showered by \PythiaEightPlot{}, \HerwigSevenPlot{}
\PythiaSixPlot{} and \HerwigSixPlot{}.

The shifts for \mwbjmax{}, without any smearing, are small and comparable
when using \HerwigSevenPlot{} or \HerwigSixPlot{}. These are not reported in
the figures, and can be obtained from the tables in the appendix.

When the smearing is applied, \HerwigSevenPlot{} and \HerwigSixPlot{} with
\bbfourl{} give comparable negative shifts, around 1~GeV.  Instead, with
\hvq{}, the displacement of the peak position (with respect to the reference
values) are around $-100 \div -200$~MeV for \HerwigSevenPlot{}, and $0\div -150$~MeV
for \HerwigSixPlot{}, for the different jet radii $R$.  Since no significant
difference between the two \Herwig{} versions was observed in the \bbfourl{}
case (where \POWHEG{} generates the hardest emission both in production and
decay), and since \hvq{} does not handle radiation in decay, this behaviour
is likely to be due to a different treatment of radiation in decay in the two
\Herwig{} versions with respect to \Pythia{}.

As for \Ebjmax{} predictions in Fig.~\ref{fig:had_showerOnly_Ebj}, we find
minor differences between \HerwigSixPlot{} and \HerwigSevenPlot{} for $R \ge
0.5$, that go in the direction to amplify the difference with respect to our
reference generator.  Similarly to \mwbjmax{}, also in this case the
discrepancies between \bbfourl{} and \hvq{} interfaced to the same shower
generator are larger for \Herwig{} than for \Pythia{}, both for the older and
newer versions.

We interpret the relative consistency of the \HerwigSevenPlot{} and
\HerwigSixPlot{} predictions with the \bbfourl{} generator as a validation of
our veto procedures and of the results presented in
Ref.~\cite{Ravasio:2018lzi}.

\section{Hadronic observables: full results}
We now summarize the results obtained by showering \hvq{} and \bbfourl{} with
the four PS programs at the full level, that is with the MPI and
hadronization switched on.  The \bbfourl{}+\PythiaSixPlot{} results shown
here and in the following sections are obtained using the {\tt FSR} veto.

\begin{figure}
  \begin{center}
  \hspace{0.5em}\includegraphics[width=0.47\textwidth]{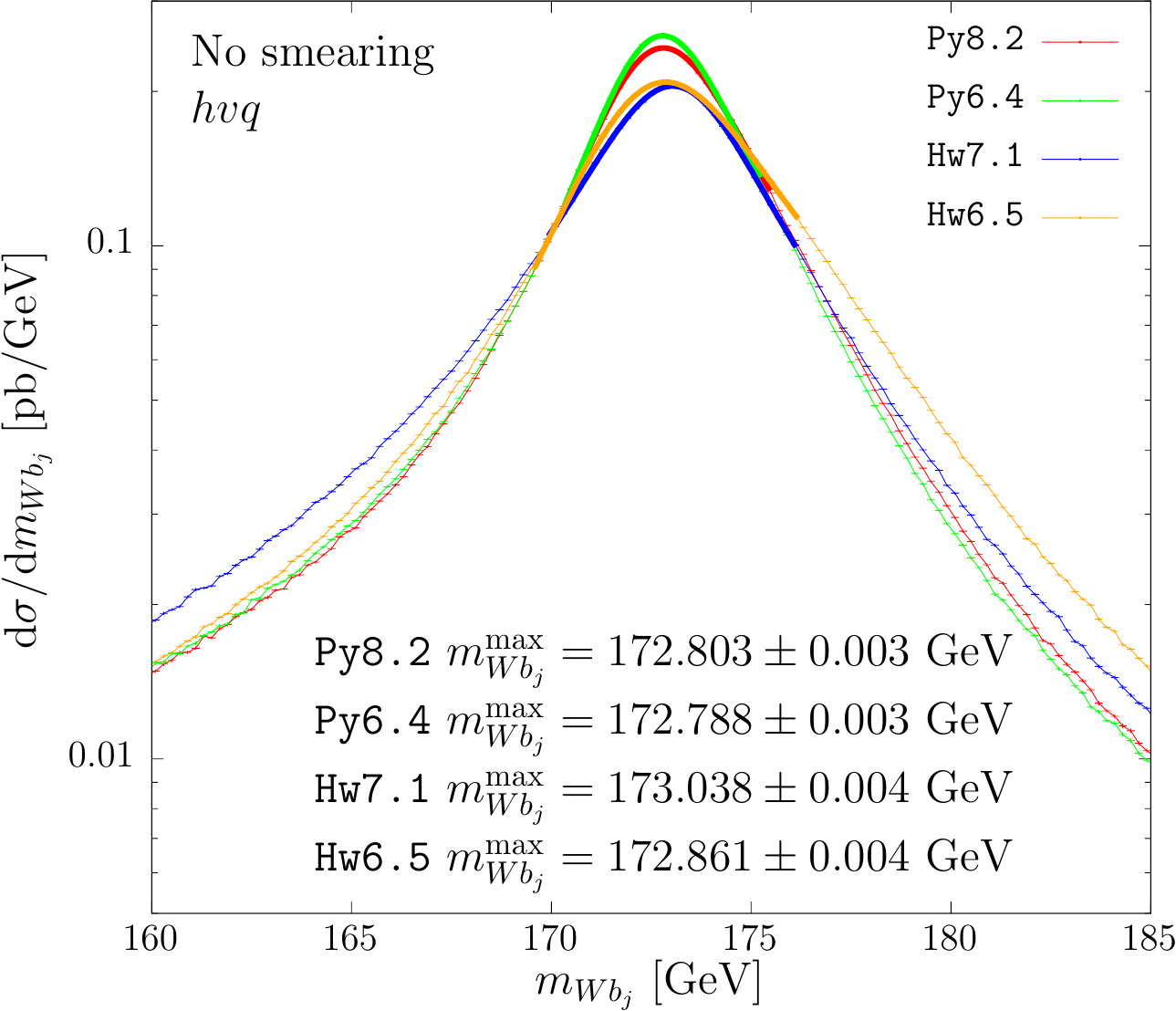} \\
  \includegraphics[width=0.48\textwidth]{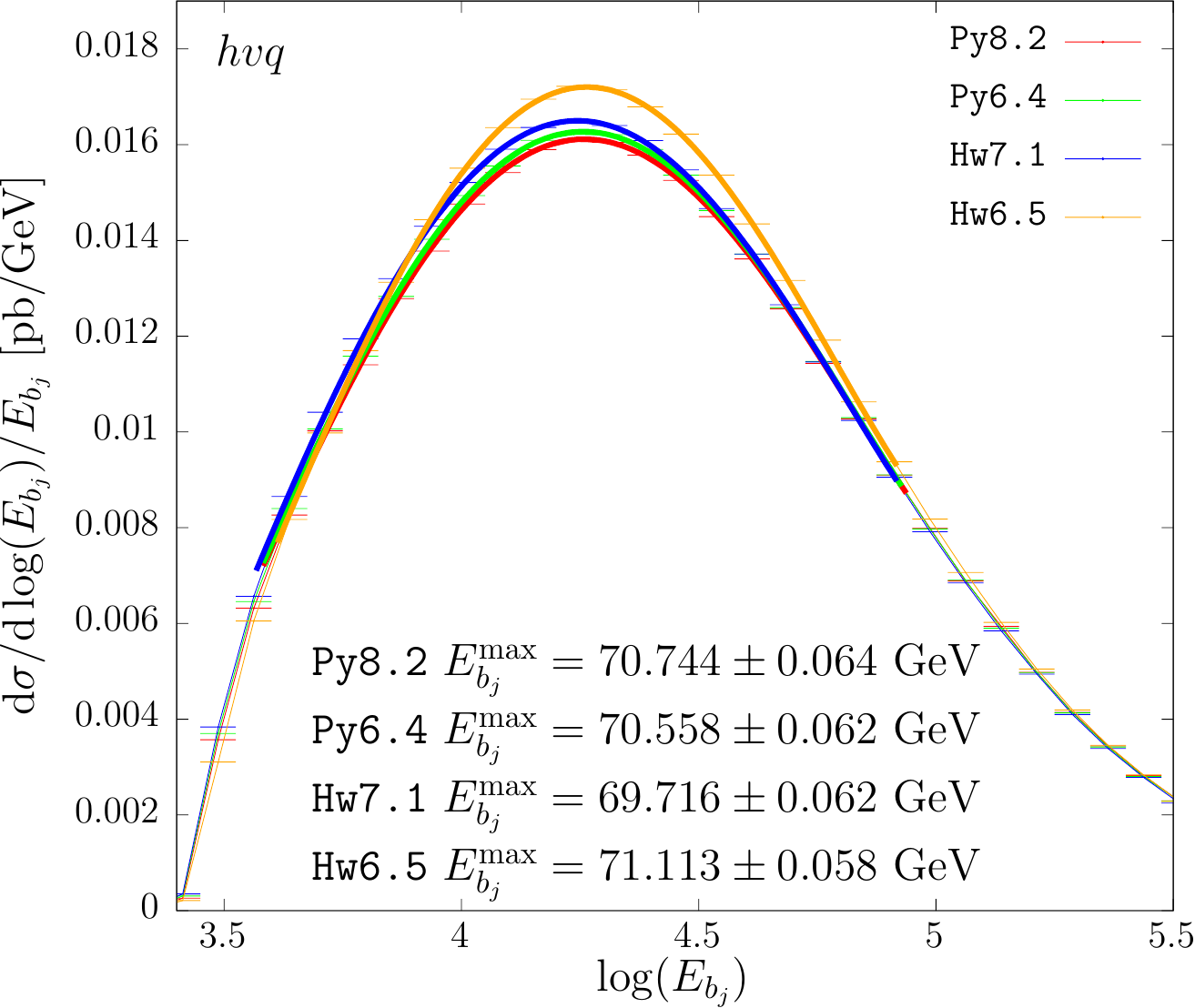}
  \end{center}
    \caption{Reconstructed-top mass~(upper pane) and $b$ jet energy
      distribution~(lower pane) obtained for the \hvq{} generator matched to
      \PythiaEightPlot{}~(red), to \PythiaSixPlot{}~(green),
      \HerwigSevenPlot{}~(blue) and \HerwigSixPlot{}~(orange). The
      hadronization and the underlying event are included.  }
    \label{fig:hvq_full}
\end{figure}

For the \hvq{} generator (see Fig.~\ref{fig:hvq_full})
we find that \PythiaSixPlot{} and \PythiaEightPlot{} yield very similar
results.  However, we find an appreciable disagreement between
\HerwigSevenPlot{} and \HerwigSixPlot{}.  We attribute it to different
implementations of MPI in the two versions of \Herwig{}, since the
predictions agreed rather well at the NLO+PS level for $R\ge
0.5$.\footnote{We stress that, among other improvements over
  \HerwigSixPlot{}, \HerwigSevenPlot{} implements a model for the treatment
  of colour reconnection.}

\begin{figure}
  \begin{center}
  \hspace{0.5em}\includegraphics[width=0.47\textwidth]{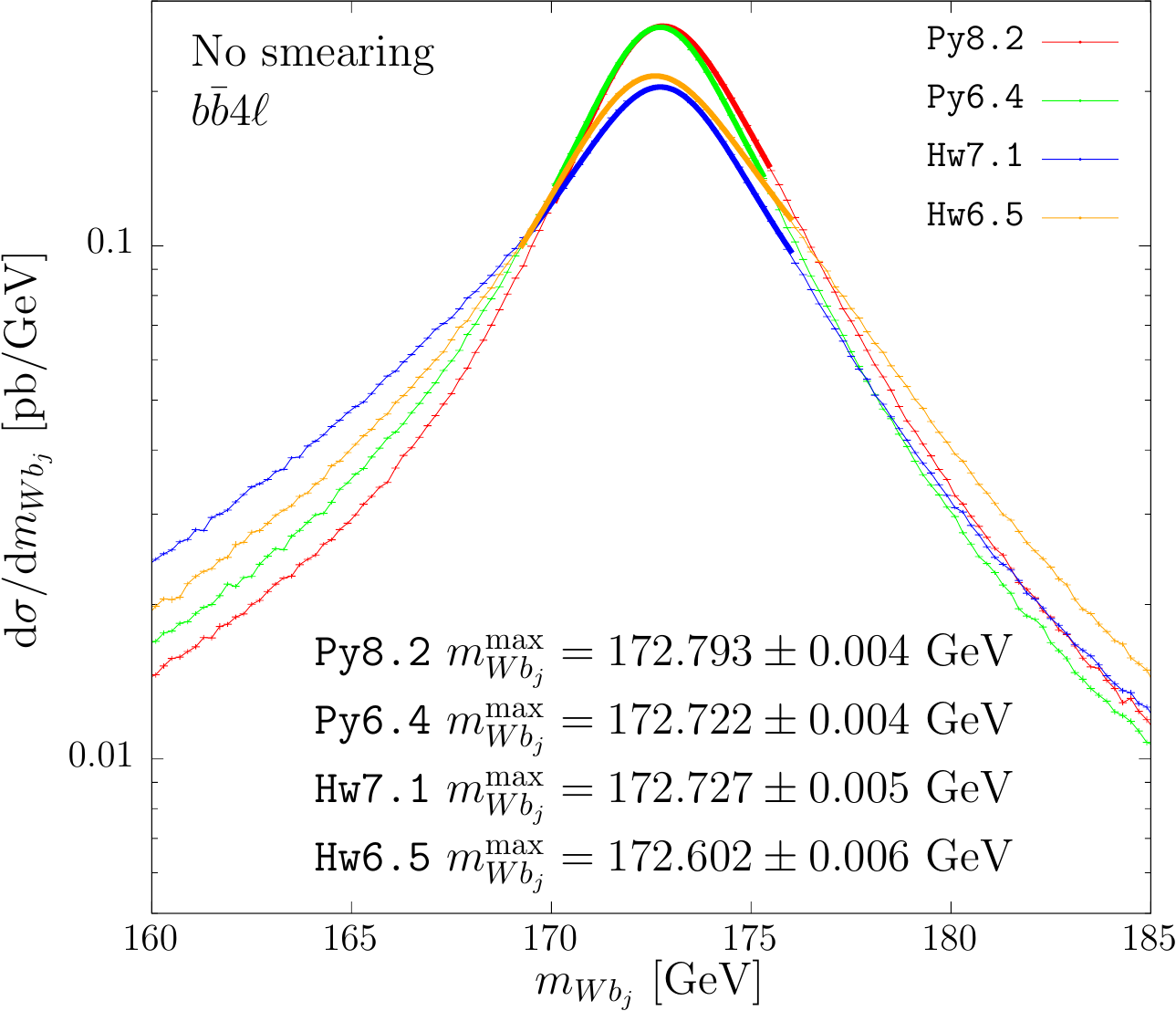} \\
  \includegraphics[width=0.48\textwidth]{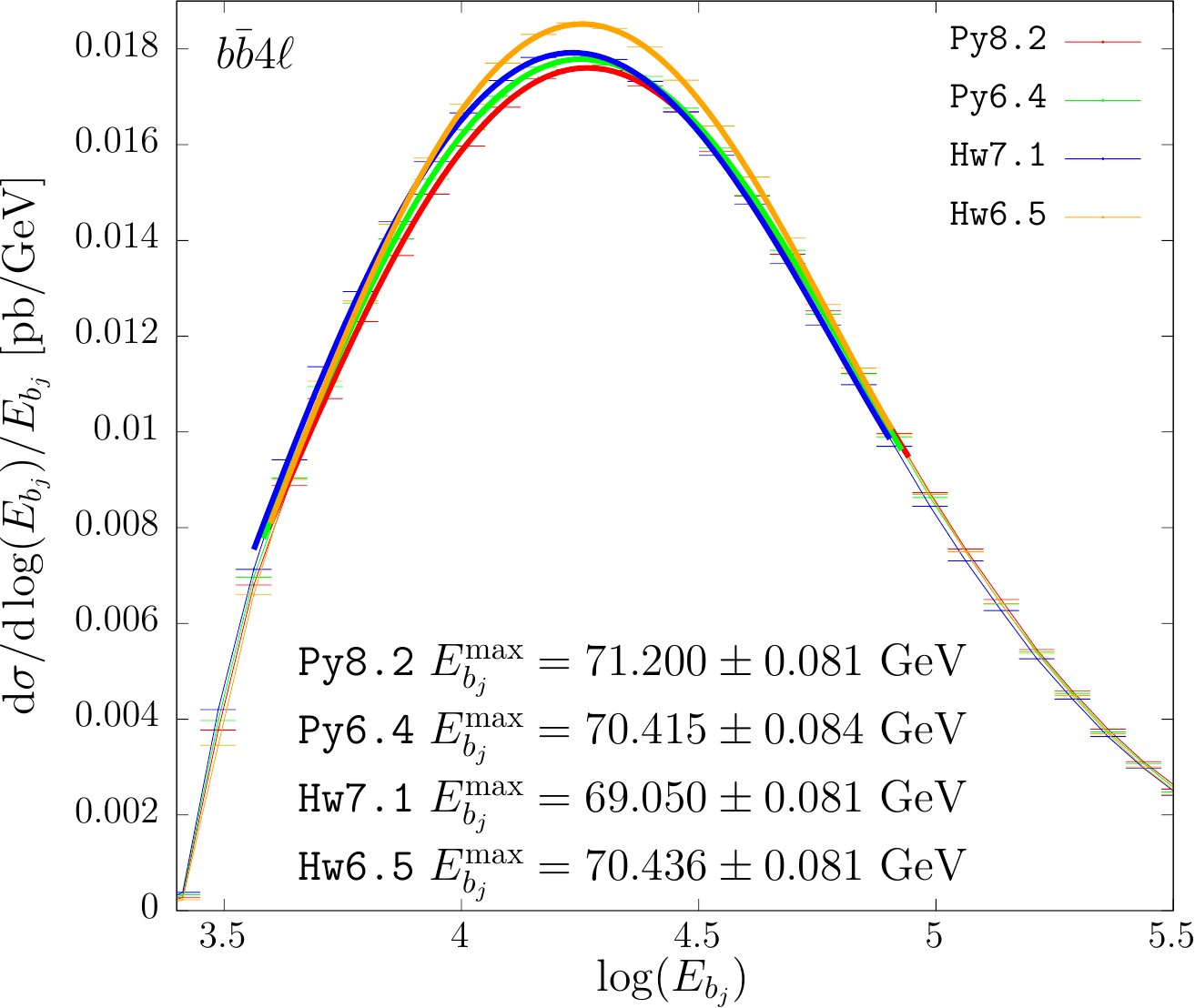}
  \end{center}
    \caption{ Reconstructed-top mass~(upper pane) and $b$ jet energy
      distribution~(lower pane) obtained for the \bbfourl{} generator matched
      to \PythiaEightPlot{}~(red), to \PythiaSixPfour{}~(green),
      \HerwigSevenPlot{}~(blue) and \HerwigSixPlot{}~(orange). The
      hadronization and the underlying event are included.  }
    \label{fig:bb4l_full}
\end{figure}

If the \bbfourl{} generator is employed (see Fig.~\ref{fig:bb4l_full}) the
same reasoning applies, but with one important difference: the discrepancy
between \PythiaEightPlot{} and \PythiaSixPlot{} is not negligible and leads
to a large \mwbjmax{} displacement when smearing is applied, similar to what
we found at the NLO+PS level.

\begin{figure}
  \centering
  \includegraphics[width=0.5\textwidth]{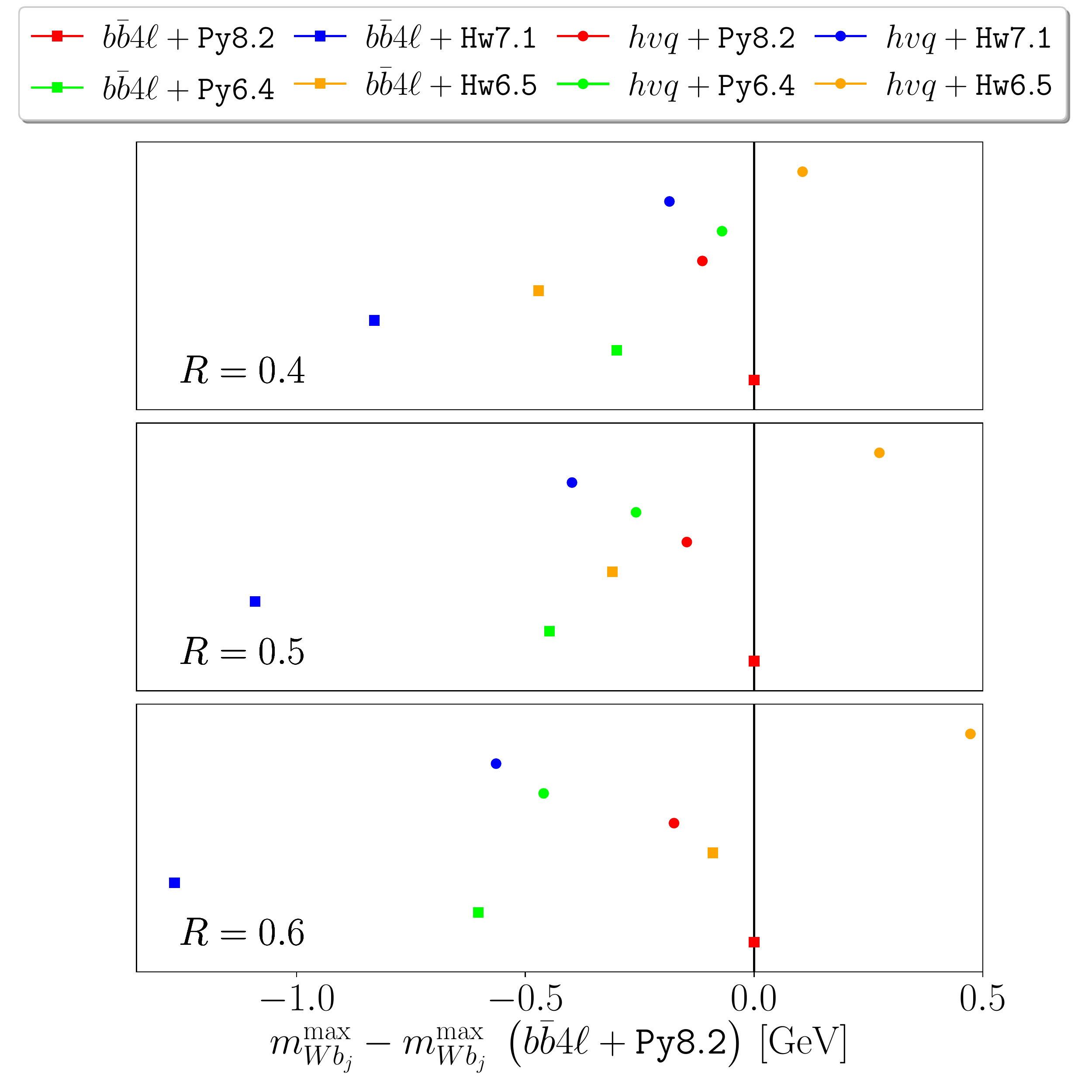}
  \caption{Results for the differences of \mwbjmax{}, including a 15~GeV
    smearing, relative to our reference generator, at the full level
    (i.e.~with the inclusion of the MPI and of the hadronization) for
    different values of jet radius $R$. The numerical values are reported in
    Tab.~\ref{tab:had_full}. The square/round dots refer to \bbfourl/\hvq{}
    results, while the colours correspond to given shower generators.}
  \label{fig:had_full_mwbj}
\end{figure}

\begin{figure}
  \centering
  \includegraphics[width=0.5\textwidth]{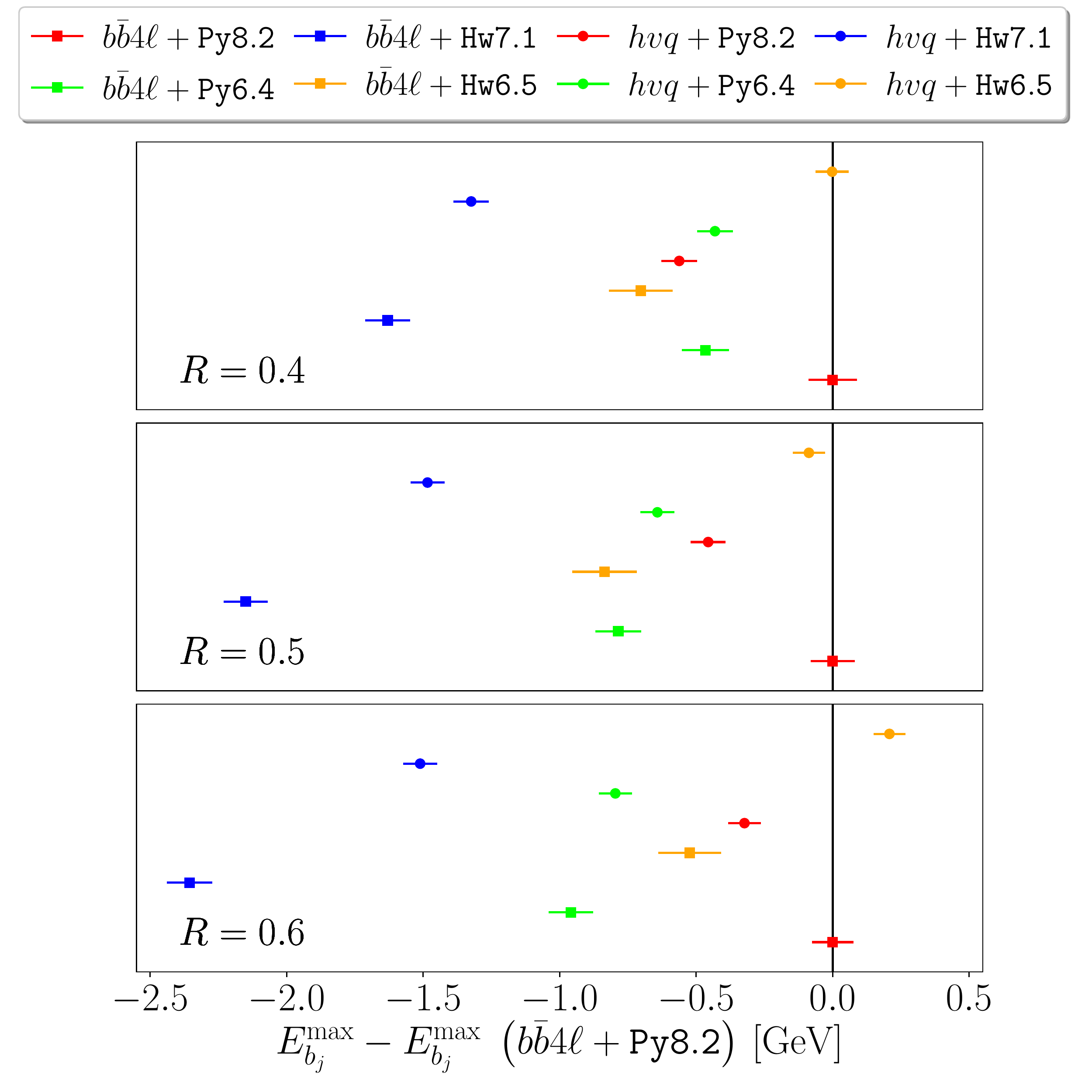}
  \caption{Same as Fig.~\ref{fig:had_full_mwbj} but for \Ebjmax.}
  \label{fig:had_full_Ebj}
\end{figure}

The \mwbj{} and \Ebj{} shifts in peak positions obtained considering several
values of the jet radius $R$, with and without smearing in the case of the
\mwbj{} distribution, are summarized in Figs.~\ref{fig:had_full_mwbj}
and~\ref{fig:had_full_Ebj}.  We notice a non-negligible $R$ dependence in the
difference between \PythiaSixPlot{} and \PythiaEightPlot{}, both in the
\hvq{} and \bbfourl{} case.  Something similar is observed in the case of
\HerwigSeven{}.  A large $R$ dependence is also observed in the case of
\HerwigSixPlot{}, but with an opposite slope when \hvq{} is used. The largest
difference with respect to our reference result is given by the
\HerwigSevenPlot, that represent a major cause of concern.  We stress that
these large differences arise in the smeared case from the mass distribution
away from the peak, i.e.~cannot be consider as an irreducible uncertainty on
the extracted mass.

Overall, we find that \bbfourl{} and \hvq{} showered with \Pythia{} exhibit
more consistency than those showered with both versions of \Herwig{}.  This
is perhaps not surprising.  Matrix-element corrections~(MEC), that have a
large impact on \hvq{} predictions (since this generator implements only LO
top decay), as implemented in the context of angular ordered parton showers
(i.e.~in \Herwig{}), are technically quite different from the way in which
the hardest top radiation is generated in \bbfourl{}, at variance with
MEC in transverse-momentum ordered showers (i.e.~in \Pythia{}).
We find that it is difficult to use this difference to dismiss
the \HerwigSevenPlot{} result, since the MEC formalism in \Herwig{}
has formally the same accuracy as the one in \Pythia{}.

\section{Leptonic observables}
The last class of observables we consider are the leptonic ones. In
Ref.~\cite{Ravasio:2018lzi} we found that these observables are only mildly
affected by non-perturbative effects (i.e.~the hadronization and the MPI),
thus we present only the results obtained at the full level and with jet
radius $R=0.5$.  However, they are likely to be strongly affected by the
parton shower, since the $W$ boson, and thus the leptons arising from its
decay, must absorb the radiation recoil to ensure four-momentum conservation.

\begin{figure*} 
  \centering
  \includegraphics[width=0.475\textwidth]{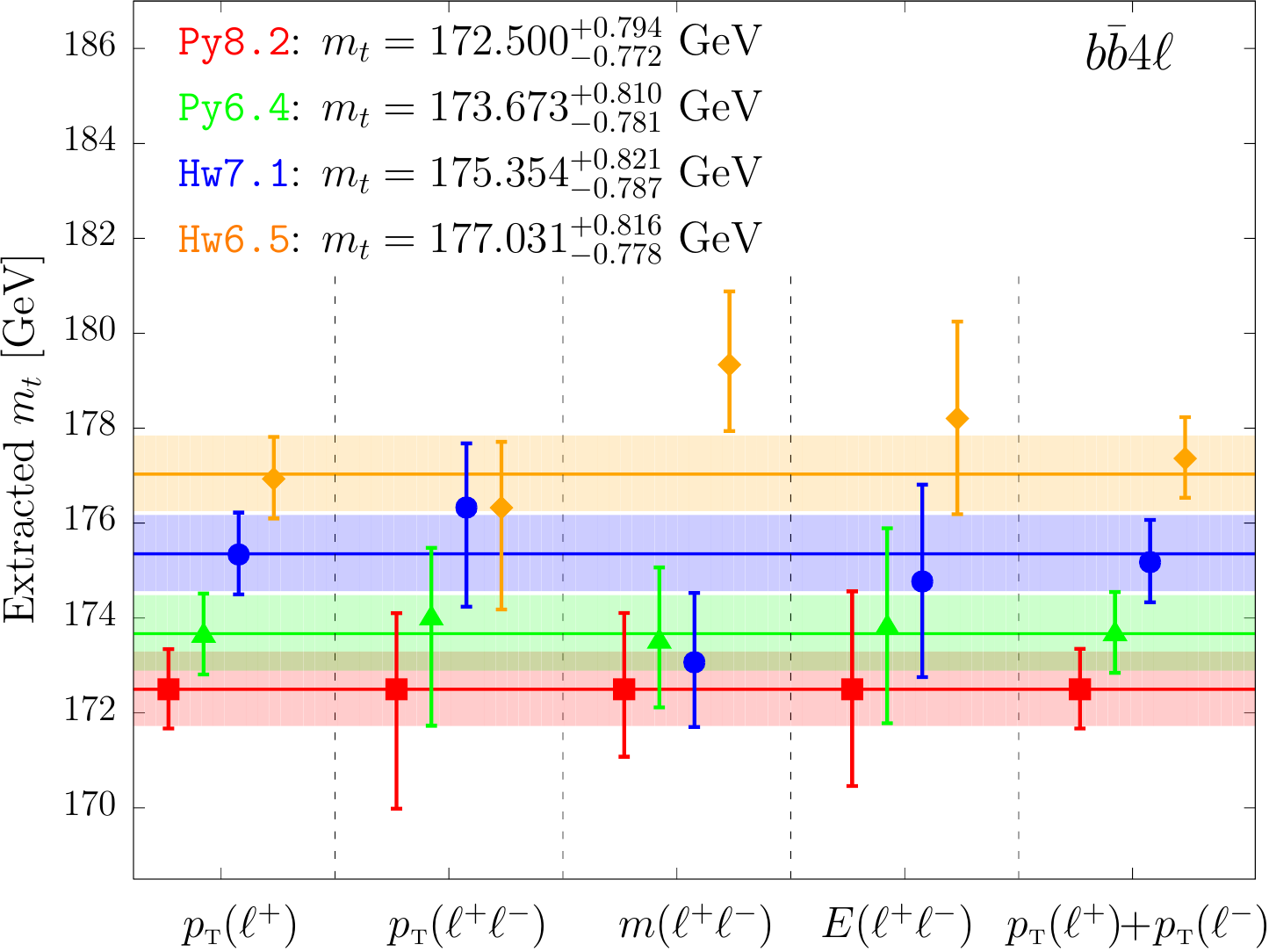}\quad
  \; \includegraphics[width=0.475\textwidth]{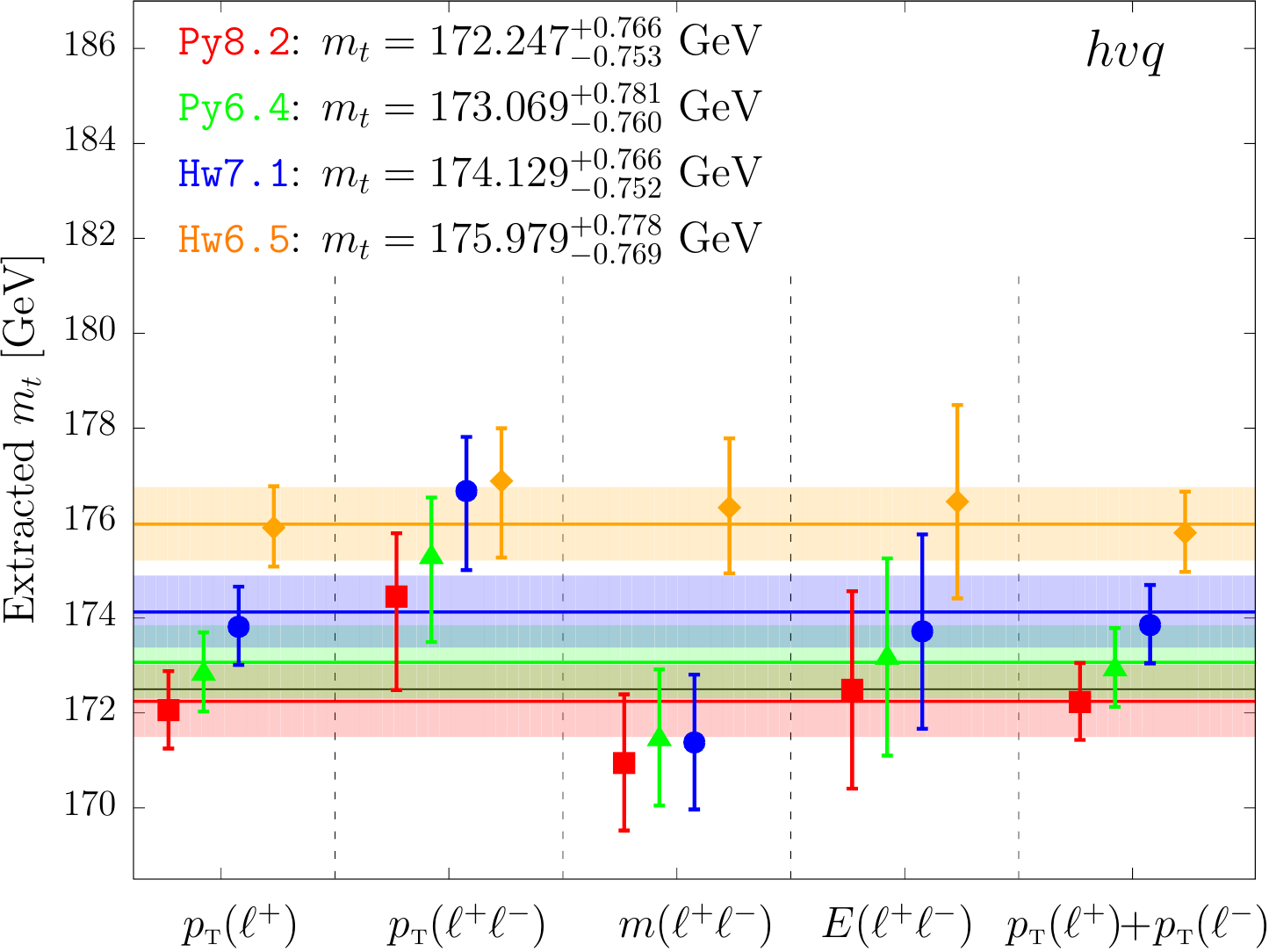}
  \caption{Extracted mass for the \bbfourl{}~(left) and \hvq~(right)
    generators matched with \PythiaEightPlot{}~(red),
    \PythiaSixPlot{}~(green), \HerwigSevenPlot{}~(blue) and
    \HerwigSixPlot{}~(orange) using the average value of the five leptonic
    observables. The horizontal band represents the weighted average of the
    results, and the black horizontal line corresponds to $\mt=172.5$~GeV,
    which is the top mass value used in the \bbfourl{}+\PythiaEightPlot{}
    reference sample.}
  \label{fig:leptons}
\end{figure*}

We extract the top mass value from the following observables:
\begin{align}
  & \langle \pt(\ell^+) \rangle, \quad \langle \pt(\ell^+\ell^-)
  \rangle,\quad \langle m(\ell^+\ell^-) \rangle,\\& \langle
  E(\ell^+\ell^-) \rangle, \quad \langle \pt(\ell^+)+
  \pt(\ell^-)\rangle. \nonumber
\end{align}
The results are presented in Tab.~\ref{tab:mass_lept}
\begin{table*}
\resizebox{1.\textwidth}{!}
  { \begin{tabular}{c||c|c|c|c||c|c|c|c||}
 \cline{2-9}
\multicolumn{1}{c|}{} &\multicolumn{4}{|c||}{$\phantom{\Big|}$ $\mt$ extracted with \bbfourl{}~[GeV]} &\multicolumn{4}{|c||}{ $\mt$ extracted with \hvq{}~[GeV]}
 \\ \cline{1-9}
\multicolumn{1}{|c||}{$\phantom{\Big|}$ observable}
& \textcolor{red}{\PythiaEightPlot{}} & \textcolor{green}{\PythiaSixPlot{}} & \textcolor{blue}{\HerwigSevenPlot{}} & \textcolor{orange}{\HerwigSixPlot{}}
& \textcolor{red}{\PythiaEightPlot{}} & \textcolor{green}{\PythiaSixPlot{}} & \textcolor{blue}{\HerwigSevenPlot{}} & \textcolor{orange}{\HerwigSixPlot{}}
 \\ \cline{1-9}
\multicolumn{1}{|c||}{$\phantom{\Big|}$$\langle \pT(\ell^+)\rangle $}
& $ 172.500_{-  0.825}^{+  0.845} $
& $ 173.649_{-  0.837}^{+  0.867} $
& $ 175.340_{-  0.841}^{+  0.884} $
& $ 176.932_{-  0.836}^{+  0.882} $
& $ 172.060_{-  0.811}^{+  0.822} $
& $ 172.847_{-  0.816}^{+  0.850} $
& $ 173.817_{-  0.803}^{+  0.843} $
& $ 175.906_{-  0.822}^{+  0.874} $
\\ \cline{1-9}
\multicolumn{1}{|c||}{$\phantom{\Big|}$$\langle \pT(\ell^+\ell^-)\rangle $}
& $ 172.500_{-  2.515}^{+  1.601} $
& $ 174.013_{-  2.282}^{+  1.466} $
& $ 176.328_{-  2.088}^{+  1.353} $
& $ 176.326_{-  2.147}^{+  1.386} $
& $ 174.451_{-  1.967}^{+  1.334} $
& $ 175.305_{-  1.809}^{+  1.236} $
& $ 176.675_{-  1.663}^{+  1.141} $
& $ 176.888_{-  1.611}^{+  1.110} $
\\ \cline{1-9}
\multicolumn{1}{|c||}{$\phantom{\Big|}$$\langle m(\ell^+\ell^-)\rangle $}
& $ 172.500_{-  1.419}^{+  1.605} $
& $ 173.523_{-  1.404}^{+  1.543} $
& $ 173.068_{-  1.363}^{+  1.459} $
& $ 179.337_{-  1.397}^{+  1.546} $
& $ 170.945_{-  1.420}^{+  1.450} $
& $ 171.472_{-  1.423}^{+  1.446} $
& $ 171.379_{-  1.412}^{+  1.429} $
& $ 176.330_{-  1.386}^{+  1.458} $
\\ \cline{1-9}
\multicolumn{1}{|c||}{$\phantom{\Big|}$$\langle E(\ell^+\ell^-)\rangle $}
& $ 172.500_{-  2.037}^{+  2.061} $
& $ 173.826_{-  2.042}^{+  2.066} $
& $ 174.771_{-  2.014}^{+  2.038} $
& $ 178.204_{-  2.017}^{+  2.040} $
& $ 172.490_{-  2.086}^{+  2.076} $
& $ 173.185_{-  2.083}^{+  2.074} $
& $ 173.720_{-  2.052}^{+  2.045} $
& $ 176.454_{-  2.039}^{+  2.034} $
\\ \cline{1-9}
\multicolumn{1}{|c||}{$\phantom{\Big|}$$\langle \pT(\ell^+)+\pT(\ell^-)\rangle $}
& $ 172.500_{-  0.827}^{+  0.852} $
& $ 173.680_{-  0.835}^{+  0.867} $
& $ 175.178_{-  0.843}^{+  0.890} $
& $ 177.362_{-  0.829}^{+  0.871} $
& $ 172.233_{-  0.802}^{+  0.821} $
& $ 172.940_{-  0.811}^{+  0.846} $
& $ 173.851_{-  0.805}^{+  0.847} $
& $ 175.794_{-  0.820}^{+  0.872} $
\\ \cline{1-9}
\multicolumn{1}{|c||}{$\phantom{\Big|}$ {\bf average}}
& $ \mathbf{172.500_{-  0.772}^{+  0.794} }$
& $ \mathbf{173.673_{-  0.781}^{+  0.810} }$
& $ \mathbf{175.354_{-  0.787}^{+  0.821} }$
& $ \mathbf{177.031_{-  0.778}^{+  0.816} }$
& $ \mathbf{172.247_{-  0.753}^{+  0.766} }$
& $ \mathbf{173.069_{-  0.760}^{+  0.781} }$
& $ \mathbf{174.129_{-  0.752}^{+  0.766} }$
& $ \mathbf{175.979_{-  0.769}^{+  0.778} }$
\\ \cline{1-9}
\end{tabular}
}
  \caption{Extracted mass for the \bbfourl{}~(left) and \hvq~(right)
    generators matched with \PythiaEightPlot{}, \PythiaSixPlot{},
    \HerwigSevenPlot{} and \HerwigSixPlot{} using the average value of the
    five leptonic observables. The average result is also shown.}
     \label{tab:mass_lept}  
\end{table*}
and their graphical display is given in Fig.~\ref{fig:leptons}.

As before, our pseudodata sample was generated with
\bbfourl{}+\PythiaEightPlot{}, and we used the other combinations of NLO+PS
generators to extract a corresponding top mass value.  We have included the
standard theoretical uncertainties as described in
Ref.~\cite{Ravasio:2018lzi}, and averaged the results obtained for the
different leptonic observables also considering the statistical correlation
among them, as suggested in Ref.~\cite{Frixione:2014ala}.

The \PythiaSixPlot{} predictions always give $m_t$ values roughly 1~GeV
larger (1.2~GeV for \bbfourl{} and 0.8~GeV for \hvq) than the corresponding
\PythiaEightPlot{} ones. This variation is of the same order of the extracted
total uncertainty on $m_t$.

The average reconstructed top mass with \HerwigSixPlot{} is nearly 2~GeV
larger than \HerwigSevenPlot{} (1.8 GeV for \bbfourl{} and 2 GeV for \hvq{}).

\section{Conclusions}
In this work we have extended the study performed in
Ref.~\cite{Ravasio:2018lzi} by also considering the \PythiaSixPlot{} and
\HerwigSixPlot{} generators.

We find that, at the NLO+PS level, the \PythiaSixPlot{} and
\PythiaEightPlot{} generators (both based upon a $\pT$-ordered shower) are
quite consistent among each other, and the same holds for \HerwigSixPlot{}
and \HerwigSevenPlot{} (both based upon an angular-ordered shower).  When
non-perturbative effects are included, we find larger differences between the
old and the new \Herwig{} versions of the PS programs, that yields a better
agreement of the old \Herwig{} version with respect to both \Pythia{}
versions (see Fig.~\ref{fig:had_full_mwbj}).

If we compare predictions for the leptonic observables, we see that the old
\Herwig{} version is further away from our reference result then the new
version.

Overall, inclusion of the older versions of the shower generators supports
what was found in Ref.~\cite{Ravasio:2018lzi}, i.e.~an indication of a large
sensitivity to the shower generator in the extraction of the top mass.

Since we have now compared four different shower and hadronization models, it
is worth asking what kind of estimate of irreducible non-perturbative
effects, potentially due to the different implementation of the shower
cut-off and the matching hadronization model. We thus consider the spread of
the \mwbjmax{} values obtained with all generators as a crude estimate of
non-perturbative effects. Looking at Tab.~\ref{tab:had_full}, the unsmeared
results from the \bbfourl{} generators, taking $R=0.5$ to avoid too large
hadronization effects (for small $R$) and too large MPI contamination (for
large $R$), we find a range from $172.601$ to $172.793$, i.e.~roughly a
200~MeV range. This result tells us that, after all, non-perturbative effects
may be well contained within presently quoted errors for direct measurements
from the experimental collaborations.

\section*{Acknowledgments}
The authors would like to acknowledge Bryan Webber for the substantial help
with the \HerwigSixPlot{\tt +Jimmy4.31} interface to \POWHEGBOX{} and the
useful discussions in the early stages of this project. The work of T.J. is
supported in part by the University of Z\"urich under the contract
K-72319-02-01 and in part by the Swiss National Science Foundation under
contract BSCGI0-157722.  P.N. acknowledges the support from Fondazione
Cariplo and Regione Lombardia, grant 2017-2070. The work of S.F.R. received
funding from the UK Science and Technology Facilities Council (grant numbers
ST/P001246/1).

\appendix

\section{Numerical results}
In this section we give the numerical results for the hadronic observables
$\mwbjmax$ and $\Ebjmax$ for both the \hvq{} and the \bbfourl{} generators,
showered with \PythiaEightPlot{}, \PythiaSixPlot{}, \HerwigSevenPlot{} and
\HerwigSixPlot{}.  In Tab.~\ref{tab:had_showerOnly} the results obtained
without the inclusion of the hadronization and MPI effects are listed. The
graphical representation of these data is given in
Figs.~\ref{fig:had_showerOnly_mwbj} and~\ref{fig:had_showerOnly_Ebj}.

The results obtained including the non-perturbative physics effects are
instead reported in Tab.~\ref{tab:had_full} and displayed in
Figs.~\ref{fig:had_full_mwbj} and~\ref{fig:had_full_Ebj}.

\begin{table*}
  \centering
  \resizebox{0.6\textwidth}{!}
  { \begin{tabular}{|c|c|c|c|c|c|}
 \cline{1-6}
 \phantom{\Big|} Obs  \phantom{\Big|} & gen & shower 
 &  \multicolumn{1}{ |c|}{ \phantom{\Big|} $R=0.4$}
 &  \multicolumn{1}{ |c|}{ \phantom{\Big|} $R=0.5$}
 &  \multicolumn{1}{ |c|}{ \phantom{\Big|} $R=0.6$} \\
 \cline{1-6}
\multirow{11}{*}{$\mwbjmax$ [GeV]}
& \multirow{5}{*}{\bbfourl{}}
 & \phantom{\Big|}{\color{red} \PythiaEightPlot{}}\phantom{\Big|}
 & $ 172.509\pm  0.002$
 & $ 172.522\pm  0.002$
 & $ 172.538\pm  0.002$
\\ \cline{3-6}
 & 
 & \phantom{\Big|}{\color{green} \PythiaSixPlot{}}\phantom{\Big|}
 & $ 172.487\pm  0.003$
 & $ 172.511\pm  0.003$
 & $ 172.538\pm  0.003$
\\ \cline{3-6}
 & 
 & \phantom{\Big|}{\color{blue} \HerwigSevenPlot{}}\phantom{\Big|}
 & $ 172.509\pm  0.002$
 & $ 172.512\pm  0.002$
 & $ 172.517\pm  0.002$
\\ \cline{3-6}
 & 
 & \phantom{\Big|}{\color{orange} \HerwigSixPlot{}}\phantom{\Big|}
 & $ 172.509\pm  0.003$
 & $ 172.515\pm  0.003$
 & $ 172.518\pm  0.003$
\\ \cline{3-6}
\cline{2-6} \\[-1.25em] \cline{2-6}
& \multirow{5}{*}{\hvq{}}
 & \phantom{\Big|}{\color{red} \PythiaEightPlot{}}\phantom{\Big|}
 & $ 172.485\pm  0.001$
 & $ 172.498\pm  0.001$
 & $ 172.513\pm  0.001$
\\ \cline{3-6}
 & 
 & \phantom{\Big|}{\color{green} \PythiaSixPlot{}}\phantom{\Big|}
 & $ 172.475\pm  0.001$
 & $ 172.499\pm  0.001$
 & $ 172.527\pm  0.001$
\\ \cline{3-6}
 & 
 & \phantom{\Big|}{\color{blue} \HerwigSevenPlot{}}\phantom{\Big|}
 & $ 172.497\pm  0.001$
 & $ 172.498\pm  0.001$
 & $ 172.499\pm  0.001$
\\ \cline{3-6}
 & 
 & \phantom{\Big|}{\color{orange} \HerwigSixPlot{}}\phantom{\Big|}
 & $ 172.495\pm  0.001$
 & $ 172.497\pm  0.001$
 & $ 172.500\pm  0.001$
\\ \cline{3-6}
\cline{2-6} \\[-1.25em] \cline{2-6}
\cline{1-6} \\[-1.25em] \cline{1-6} 
\multirow{11}{1.5cm}{\centering $\mwbjmax$ [GeV]  smearing}
& \multirow{5}{*}{\bbfourl{}}
 & \phantom{\Big|}{\color{red} \PythiaEightPlot{}}\phantom{\Big|}
 & $ 170.569\pm  0.002$
 & $ 171.403\pm  0.002$
 & $ 172.117\pm  0.002$
\\ \cline{3-6}
 & 
 & \phantom{\Big|}{\color{green} \PythiaSixPlot{}}\phantom{\Big|}
 & $ 170.273\pm  0.002$
 & $ 171.117\pm  0.002$
 & $ 171.859\pm  0.002$
\\ \cline{3-6}
 & 
 & \phantom{\Big|}{\color{blue} \HerwigSevenPlot{}}\phantom{\Big|}
 & $ 169.699\pm  0.002$
 & $ 170.419\pm  0.002$
 & $ 171.108\pm  0.002$
\\ \cline{3-6}
 & 
 & \phantom{\Big|}{\color{orange} \HerwigSixPlot{}}\phantom{\Big|}
 & $ 169.631\pm  0.002$
 & $ 170.451\pm  0.002$
 & $ 171.223\pm  0.002$
\\ \cline{3-6}
\cline{2-6}\\[-1.25em] \cline{2-6}
& \multirow{5}{*}{\hvq{}}
 & \phantom{\Big|}{\color{red} \PythiaEightPlot{}}\phantom{\Big|}
 & $ 170.518\pm  0.001$
 & $ 171.315\pm  0.001$
 & $ 171.996\pm  0.001$
\\ \cline{3-6}
 & 
 & \phantom{\Big|}{\color{green} \PythiaSixPlot{}}\phantom{\Big|}
 & $ 170.594\pm  0.001$
 & $ 171.384\pm  0.001$
 & $ 172.064\pm  0.001$
\\ \cline{3-6}
 & 
 & \phantom{\Big|}{\color{blue} \HerwigSevenPlot{}}\phantom{\Big|}
 & $ 170.464\pm  0.001$
 & $ 171.202\pm  0.001$
 & $ 171.867\pm  0.001$
\\ \cline{3-6}
 & 
 & \phantom{\Big|}{\color{orange} \HerwigSixPlot{}}\phantom{\Big|}
 & $ 170.560\pm  0.001$
 & $ 171.283\pm  0.001$
 & $ 171.953\pm  0.001$
\\ \cline{3-6}
\cline{2-6}\\[-1.25em] \cline{2-6}
\cline{1-6}\\[-1.25em] \cline{1-6}
\multirow{11}{*}{$\Ebjmax$ [GeV]}
& \multirow{5}{*}{\bbfourl{}}
 & \phantom{\Big|}{\color{red} \PythiaEightPlot{}}\phantom{\Big|}
 & $  67.145\pm  0.086$
 & $  69.614\pm  0.082$
 & $  71.747\pm  0.080$
\\ \cline{3-6}
 & 
 & \phantom{\Big|}{\color{green} \PythiaSixPlot{}}\phantom{\Big|}
 & $  66.709\pm  0.123$
 & $  69.042\pm  0.120$
 & $  71.181\pm  0.117$
\\ \cline{3-6}
 & 
 & \phantom{\Big|}{\color{blue} \HerwigSevenPlot{}}\phantom{\Big|}
 & $  65.847\pm  0.084$
 & $  67.948\pm  0.083$
 & $  69.945\pm  0.082$
\\ \cline{3-6}
 & 
 & \phantom{\Big|}{\color{orange} \HerwigSixPlot{}}\phantom{\Big|}
 & $  65.562\pm  0.121$
 & $  67.631\pm  0.121$
 & $  69.809\pm  0.114$
\\ \cline{3-6}
\cline{2-6}\\[-1.25em] \cline{2-6}
& \multirow{5}{*}{\hvq{}}
 & \phantom{\Big|}{\color{red} \PythiaEightPlot{}}\phantom{\Big|}
 & $  66.791\pm  0.068$
 & $  69.357\pm  0.063$
 & $  71.598\pm  0.061$
\\ \cline{3-6}
 & 
 & \phantom{\Big|}{\color{green} \PythiaSixPlot{}}\phantom{\Big|}
 & $  66.768\pm  0.067$
 & $  69.257\pm  0.065$
 & $  71.465\pm  0.062$
\\ \cline{3-6}
 & 
 & \phantom{\Big|}{\color{blue} \HerwigSevenPlot{}}\phantom{\Big|}
 & $  66.276\pm  0.065$
 & $  68.650\pm  0.063$
 & $  70.819\pm  0.061$
\\ \cline{3-6}
 & 
 & \phantom{\Big|}{\color{orange} \HerwigSixPlot{}}\phantom{\Big|}
 & $  66.699\pm  0.061$
 & $  68.923\pm  0.060$
 & $  71.000\pm  0.057$
\\ \cline{3-6}
\cline{1-6}
\end{tabular}
}
  \caption{Results for \mwbjmax{} and \Ebjmax{} at the NLO+PS level, showered
    by \Pythia{} and \Herwig{}, without hadronization or MPI effects, for
    different values of jet radius $R$.}
  \label{tab:had_showerOnly}
\end{table*}
\begin{table*}
  \centering
  \resizebox{0.60\textwidth}{!}
  { \begin{tabular}{|c|c|c|c|c|c|}
 \cline{1-6}
 \phantom{\Big|} Obs  \phantom{\Big|} & gen & shower 
 &  \multicolumn{1}{ |c|}{ \phantom{\Big|} $R=0.4$}
 &  \multicolumn{1}{ |c|}{ \phantom{\Big|} $R=0.5$}
 &  \multicolumn{1}{ |c|}{ \phantom{\Big|} $R=0.6$} \\
 \cline{1-6}
\multirow{11}{*}{$\mwbjmax$ [GeV]}
& \multirow{5}{*}{\bbfourl{}}
 & \phantom{\Big|}{\color{red} \PythiaEightPlot{}}\phantom{\Big|}
 & $ 172.156\pm  0.004$
 & $ 172.793\pm  0.004$
 & $ 173.436\pm  0.005$
\\ \cline{3-6}
 & 
 & \phantom{\Big|}{\color{green} \PythiaSixPlot{}}\phantom{\Big|}
 & $ 172.191\pm  0.006$
 & $ 172.723\pm  0.006$
 & $ 173.252\pm  0.007$
\\ \cline{3-6}
 & 
 & \phantom{\Big|}{\color{blue} \HerwigSevenPlot{}}\phantom{\Big|}
 & $ 172.253\pm  0.005$
 & $ 172.727\pm  0.005$
 & $ 173.183\pm  0.006$
\\ \cline{3-6}
 & 
 & \phantom{\Big|}{\color{orange} \HerwigSixPlot{}}\phantom{\Big|}
 & $ 171.985\pm  0.008$
 & $ 172.601\pm  0.008$
 & $ 173.175\pm  0.009$
\\ \cline{3-6}
\cline{2-6} \\[-1.25em] \cline{2-6}
& \multirow{5}{*}{\hvq{}}
 & \phantom{\Big|}{\color{red} \PythiaEightPlot{}}\phantom{\Big|}
 & $ 172.203\pm  0.003$
 & $ 172.803\pm  0.003$
 & $ 173.429\pm  0.004$
\\ \cline{3-6}
 & 
 & \phantom{\Big|}{\color{green} \PythiaSixPlot{}}\phantom{\Big|}
 & $ 172.274\pm  0.003$
 & $ 172.788\pm  0.003$
 & $ 173.270\pm  0.004$
\\ \cline{3-6}
 & 
 & \phantom{\Big|}{\color{blue} \HerwigSevenPlot{}}\phantom{\Big|}
 & $ 172.573\pm  0.004$
 & $ 173.038\pm  0.004$
 & $ 173.460\pm  0.004$
\\ \cline{3-6}
 & 
 & \phantom{\Big|}{\color{orange} \HerwigSixPlot{}}\phantom{\Big|}
 & $ 172.224\pm  0.004$
 & $ 172.861\pm  0.004$
 & $ 173.419\pm  0.005$
\\ \cline{3-6}
\cline{2-6} \\[-1.25em] \cline{2-6}
\cline{1-6} \\[-1.25em] \cline{1-6} 
\multirow{11}{1.5cm}{\centering $\mwbjmax$ [GeV]  smearing}
& \multirow{5}{*}{\bbfourl{}}
 & \phantom{\Big|}{\color{red} \PythiaEightPlot{}}\phantom{\Big|}
 & $ 171.018\pm  0.002$
 & $ 172.717\pm  0.002$
 & $ 174.378\pm  0.002$
\\ \cline{3-6}
 & 
 & \phantom{\Big|}{\color{green} \PythiaSixPlot{}}\phantom{\Big|}
 & $ 170.716\pm  0.002$
 & $ 172.267\pm  0.002$
 & $ 173.774\pm  0.002$
\\ \cline{3-6}
 & 
 & \phantom{\Big|}{\color{blue} \HerwigSevenPlot{}}\phantom{\Big|}
 & $ 170.188\pm  0.002$
 & $ 171.626\pm  0.002$
 & $ 173.111\pm  0.002$
\\ \cline{3-6}
 & 
 & \phantom{\Big|}{\color{orange} \HerwigSixPlot{}}\phantom{\Big|}
 & $ 170.547\pm  0.002$
 & $ 172.407\pm  0.002$
 & $ 174.288\pm  0.003$
\\ \cline{3-6}
\cline{2-6}\\[-1.25em] \cline{2-6}
& \multirow{5}{*}{\hvq{}}
 & \phantom{\Big|}{\color{red} \PythiaEightPlot{}}\phantom{\Big|}
 & $ 170.905\pm  0.001$
 & $ 172.570\pm  0.001$
 & $ 174.203\pm  0.001$
\\ \cline{3-6}
 & 
 & \phantom{\Big|}{\color{green} \PythiaSixPlot{}}\phantom{\Big|}
 & $ 170.948\pm  0.001$
 & $ 172.459\pm  0.001$
 & $ 173.918\pm  0.001$
\\ \cline{3-6}
 & 
 & \phantom{\Big|}{\color{blue} \HerwigSevenPlot{}}\phantom{\Big|}
 & $ 170.833\pm  0.001$
 & $ 172.319\pm  0.001$
 & $ 173.814\pm  0.001$
\\ \cline{3-6}
 & 
 & \phantom{\Big|}{\color{orange} \HerwigSixPlot{}}\phantom{\Big|}
 & $ 171.124\pm  0.001$
 & $ 172.991\pm  0.001$
 & $ 174.851\pm  0.001$
\\ \cline{3-6}
\cline{2-6}\\[-1.25em] \cline{2-6}
\cline{1-6}\\[-1.25em] \cline{1-6}
\multirow{11}{*}{$\Ebjmax$ [GeV]}
& \multirow{5}{*}{\bbfourl{}}
 & \phantom{\Big|}{\color{red} \PythiaEightPlot{}}\phantom{\Big|}
 & $  67.792\pm  0.089$
 & $  71.200\pm  0.081$
 & $  74.454\pm  0.076$
\\ \cline{3-6}
 & 
 & \phantom{\Big|}{\color{green} \PythiaSixPlot{}}\phantom{\Big|}
 & $  67.205\pm  0.123$
 & $  70.343\pm  0.117$
 & $  73.420\pm  0.113$
\\ \cline{3-6}
 & 
 & \phantom{\Big|}{\color{blue} \HerwigSevenPlot{}}\phantom{\Big|}
 & $  66.162\pm  0.083$
 & $  69.050\pm  0.081$
 & $  72.098\pm  0.083$
\\ \cline{3-6}
 & 
 & \phantom{\Big|}{\color{orange} \HerwigSixPlot{}}\phantom{\Big|}
 & $  67.089\pm  0.117$
 & $  70.364\pm  0.118$
 & $  73.930\pm  0.115$
\\ \cline{3-6}
\cline{2-6}\\[-1.25em] \cline{2-6}
& \multirow{5}{*}{\hvq{}}
 & \phantom{\Big|}{\color{red} \PythiaEightPlot{}}\phantom{\Big|}
 & $  67.230\pm  0.066$
 & $  70.744\pm  0.064$
 & $  74.131\pm  0.060$
\\ \cline{3-6}
 & 
 & \phantom{\Big|}{\color{green} \PythiaSixPlot{}}\phantom{\Big|}
 & $  67.361\pm  0.066$
 & $  70.558\pm  0.062$
 & $  73.658\pm  0.061$
\\ \cline{3-6}
 & 
 & \phantom{\Big|}{\color{blue} \HerwigSevenPlot{}}\phantom{\Big|}
 & $  66.468\pm  0.065$
 & $  69.716\pm  0.062$
 & $  72.943\pm  0.062$
\\ \cline{3-6}
 & 
 & \phantom{\Big|}{\color{orange} \HerwigSixPlot{}}\phantom{\Big|}
 & $  67.790\pm  0.060$
 & $  71.113\pm  0.058$
 & $  74.622\pm  0.057$
\\ \cline{3-6}
\cline{1-6}
\end{tabular}
}
  \caption{\mwbjmax{} and \Ebjmax{} results at the full level, i.e.~with the
    inclusion of the MPI and of the hadronization.}
  \label{tab:had_full}
\end{table*}

\clearpage


\begin{thebibliography}{10}

\bibitem{Ravasio:2018lzi}
S.~Ferrario~Ravasio, T.~Ježo, P.~Nason and C.~Oleari, \emph{{A Theoretical
  Study of Top-Mass Measurements at the LHC Using NLO+PS Generators of
  Increasing Accuracy}},
  \href{http://dx.doi.org/10.1140/epjc/s10052-018-5909-7}{\emph{Eur. Phys. J.}
  {\bf C78} (2018) 458}, [\href{https://arxiv.org/abs/1801.03944}{{\tt
  1801.03944}}].

\bibitem{Frixione:2007nw}
S.~Frixione, P.~Nason and G.~Ridolfi, \emph{{A Positive-weight
  next-to-leading-order Monte Carlo for heavy flavour hadroproduction}},
  \href{http://dx.doi.org/10.1088/1126-6708/2007/09/126}{\emph{JHEP} {\bf 09}
  (2007) 126}, [\href{https://arxiv.org/abs/0707.3088}{{\tt 0707.3088}}].

\bibitem{Campbell:2014kua}
J.~M. Campbell, R.~K. Ellis, P.~Nason and E.~Re, \emph{{Top-pair production and
  decay at NLO matched with parton showers}},
  \href{http://dx.doi.org/10.1007/JHEP04(2015)114}{\emph{JHEP} {\bf 04} (2015)
  114}, [\href{https://arxiv.org/abs/1412.1828}{{\tt 1412.1828}}].

\bibitem{Jezo:2016ujg}
T.~Je\v{z}o, J.~M. Lindert, P.~Nason, C.~Oleari and S.~Pozzorini, \emph{{An
  NLO+PS generator for ${{t \bar{t}}}$ and ${{W t}}$ production and decay
  including non-resonant and interference effects}},
  \href{http://dx.doi.org/10.1140/epjc/s10052-016-4538-2}{\emph{Eur. Phys. J.}
  {\bf C76} (2016) 691}, [\href{https://arxiv.org/abs/1607.04538}{{\tt
  1607.04538}}].

\bibitem{Nason:2004rx}
P.~Nason, \emph{{A New method for combining NLO QCD with shower Monte Carlo
  algorithms}},
  \href{http://dx.doi.org/10.1088/1126-6708/2004/11/040}{\emph{JHEP} {\bf 11}
  (2004) 040}, [\href{https://arxiv.org/abs/hep-ph/0409146}{{\tt
  hep-ph/0409146}}].

\bibitem{Frixione:2007vw}
S.~Frixione, P.~Nason and C.~Oleari, \emph{{Matching NLO QCD computations with
  Parton Shower simulations: the POWHEG method}},
  \href{http://dx.doi.org/10.1088/1126-6708/2007/11/070}{\emph{JHEP} {\bf 11}
  (2007) 070}, [\href{https://arxiv.org/abs/0709.2092}{{\tt 0709.2092}}].

\bibitem{Alioli:2010xd}
S.~Alioli, P.~Nason, C.~Oleari and E.~Re, \emph{{A general framework for
  implementing NLO calculations in shower Monte Carlo programs: the POWHEG
  BOX}}, \href{http://dx.doi.org/10.1007/JHEP06(2010)043}{\emph{JHEP} {\bf 06}
  (2010) 043}, [\href{https://arxiv.org/abs/1002.2581}{{\tt 1002.2581}}].

\bibitem{Jezo:2015aia}
T.~Je\v{z}o and P.~Nason, \emph{{On the Treatment of Resonances in
  Next-to-Leading Order Calculations Matched to a Parton Shower}},
  \href{http://dx.doi.org/10.1007/JHEP12(2015)065}{\emph{JHEP} {\bf 12} (2015)
  065}, [\href{https://arxiv.org/abs/1509.09071}{{\tt 1509.09071}}].

\bibitem{Sjostrand:2014zea}
T.~Sjöstrand, S.~Ask, J.~R. Christiansen, R.~Corke, N.~Desai, P.~Ilten et~al.,
  \emph{{An Introduction to PYTHIA 8.2}},
  \href{http://dx.doi.org/10.1016/j.cpc.2015.01.024}{\emph{Comput. Phys.
  Commun.} {\bf 191} (2015) 159--177},
  [\href{https://arxiv.org/abs/1410.3012}{{\tt 1410.3012}}].

\bibitem{Bahr:2008pv}
M.~Bahr et~al., \emph{{Herwig++ Physics and Manual}},
  \href{http://dx.doi.org/10.1140/epjc/s10052-008-0798-9}{\emph{Eur. Phys. J.}
  {\bf C58} (2008) 639--707}, [\href{https://arxiv.org/abs/0803.0883}{{\tt
  0803.0883}}].

\bibitem{Bellm:2015jjp}
J.~Bellm et~al., \emph{{Herwig 7.0/Herwig++ 3.0 release note}},
  \href{http://dx.doi.org/10.1140/epjc/s10052-016-4018-8}{\emph{Eur. Phys. J.}
  {\bf C76} (2016) 196}, [\href{https://arxiv.org/abs/1512.01178}{{\tt
  1512.01178}}].

\bibitem{Agashe:2016bok}
K.~Agashe, R.~Franceschini, D.~Kim and M.~Schulze, \emph{{Top quark mass
  determination from the energy peaks of b-jets and B-hadrons at NLO QCD}},
  \href{http://dx.doi.org/10.1140/epjc/s10052-016-4494-x}{\emph{Eur. Phys. J.}
  {\bf C76} (2016) 636}, [\href{https://arxiv.org/abs/1603.03445}{{\tt
  1603.03445}}].

\bibitem{Frixione:2014ala}
S.~Frixione and A.~Mitov, \emph{{Determination of the top quark mass from
  leptonic observables}},
  \href{http://dx.doi.org/10.1007/JHEP09(2014)012}{\emph{JHEP} {\bf 09} (2014)
  012}, [\href{https://arxiv.org/abs/1407.2763}{{\tt 1407.2763}}].

\bibitem{Sjostrand:2006za}
T.~Sjostrand, S.~Mrenna and P.~Z. Skands, \emph{{PYTHIA 6.4 Physics and
  Manual}}, \href{http://dx.doi.org/10.1088/1126-6708/2006/05/026}{\emph{JHEP}
  {\bf 05} (2006) 026}, [\href{https://arxiv.org/abs/hep-ph/0603175}{{\tt
  hep-ph/0603175}}].

\bibitem{Corcella:2000bw}
G.~Corcella, I.~G. Knowles, G.~Marchesini, S.~Moretti, K.~Odagiri,
  P.~Richardson et~al., \emph{{HERWIG 6: An Event generator for hadron emission
  reactions with interfering gluons (including supersymmetric processes)}},
  \href{http://dx.doi.org/10.1088/1126-6708/2001/01/010}{\emph{JHEP} {\bf 01}
  (2001) 010}, [\href{https://arxiv.org/abs/hep-ph/0011363}{{\tt
  hep-ph/0011363}}].

\bibitem{Butterworth:1996zw}
J.~M. Butterworth, J.~R. Forshaw and M.~H. Seymour, \emph{{Multiparton
  interactions in photoproduction at HERA}},
  \href{http://dx.doi.org/10.1007/BF02909195, 10.1007/s002880050286}{\emph{Z.
  Phys.} {\bf C72} (1996) 637--646},
  [\href{https://arxiv.org/abs/hep-ph/9601371}{{\tt hep-ph/9601371}}].

\bibitem{Gieseke:2003rz}
S.~Gieseke, P.~Stephens and B.~Webber, \emph{{New formalism for QCD parton
  showers}}, \href{http://dx.doi.org/10.1088/1126-6708/2003/12/045}{\emph{JHEP}
  {\bf 12} (2003) 045}, [\href{https://arxiv.org/abs/hep-ph/0310083}{{\tt
  hep-ph/0310083}}].

\bibitem{Bahr:2008dy}
M.~Bahr, S.~Gieseke and M.~H. Seymour, \emph{{Simulation of multiple partonic
  interactions in Herwig++}},
  \href{http://dx.doi.org/10.1088/1126-6708/2008/07/076}{\emph{JHEP} {\bf 07}
  (2008) 076}, [\href{https://arxiv.org/abs/0803.3633}{{\tt 0803.3633}}].

\bibitem{Gieseke:2012ft}
S.~Gieseke, C.~Rohr and A.~Siodmok, \emph{{Colour reconnections in Herwig++}},
  \href{http://dx.doi.org/10.1140/epjc/s10052-012-2225-5}{\emph{Eur. Phys. J.}
  {\bf C72} (2012) 2225}, [\href{https://arxiv.org/abs/1206.0041}{{\tt
  1206.0041}}].

\bibitem{Skands:2010ak}
P.~Z. Skands, \emph{{Tuning Monte Carlo Generators: The Perugia Tunes}},
  \href{http://dx.doi.org/10.1103/PhysRevD.82.074018}{\emph{Phys. Rev.} {\bf
  D82} (2010) 074018}, [\href{https://arxiv.org/abs/1005.3457}{{\tt
  1005.3457}}].

\bibitem{Boos:2001cv}
E.~Boos et~al., \emph{{Generic user process interface for event generators}},
  in \emph{{Physics at TeV colliders. Proceedings, Euro Summer School, Les
  Houches, France, May 21-June 1, 2001}}, 2001.
\newblock \href{https://arxiv.org/abs/hep-ph/0109068}{{\tt hep-ph/0109068}}.

\bibitem{ATLAS:2011gmi}
{\scshape ATLAS} collaboration, \emph{{New ATLAS event generator tunes to 2010
  data}}, .

\end{thebibliography}
\providecommand{\href}[2]{#2}\begingroup\raggedright\endgroup

\end{document}